\documentclass[%
 reprint,
superscriptaddress,
 prb,
]{revtex4-2}
\usepackage{graphicx}% Include figure files
\usepackage{dcolumn}% Align table columns on decimal point
\usepackage{bm}% bold math
\usepackage{float}
\usepackage{array}
\usepackage{booktabs}
\usepackage{amssymb}
\usepackage{bbding}
\usepackage[dvipsnames]{xcolor}

\usepackage[colorlinks=true,linkcolor=blue,anchorcolor=red,citecolor=blue,urlcolor=blue]{hyperref}

\usepackage[mathlines]{lineno}% Enable numbering of text and display math
\begin{document}
\title{All-electrically controlled spintronics in altermagnetic heterostructures}

\author{Pei-Hao Fu}
\email{phy.phfu@gmail.com}
\affiliation{School of Science, Sun Yat-sen University, Shenzhen 518107, China}
\affiliation{School of Physics and Materials Science, Guangzhou University, Guangzhou 510006, China}

\author{Qianqian Lv}
%\email{2024000090@sziit.edu.cn}
\affiliation{School of Humanities and Basic Sciences, Shenzhen Institute of Information Technology, Shenzhen 518172, China}

\author{Yong Xu}
%\email{xuyong_nbut@163.com}
\affiliation{Institute of Materials, Ningbo University of Technology, Ningbo 315016, China}

\author{Jorge Cayao}
%\email{jorge.cayao@physics.uu.se}
\affiliation{Department of Physics and Astronomy, Uppsala University, Box 516, S-751 20 Uppsala, Sweden}

\author{Jun-Feng Liu}
\email{phjfliu@gzhu.edu.cn}
\affiliation{School of Physics and Materials Science, Guangzhou University, Guangzhou 510006, China}

\author{Xiang-Long Yu}
\email{yuxlong6@mail.sysu.edu.cn}
\affiliation{School of Science, Sun Yat-sen University, Shenzhen 518107, China}

%%%%%%%%%%%%%%%%%%%%%%%%%%%%%%%%%%%%%%%%%%%%%%%%%%%%%%%%%%%%%%%%%%%%%%
\begin{abstract}
The recent discovery of altermagnets, which exhibit spin splitting without net magnetization, opens new directions for spintronics beyond the limits of ferromagnets, antiferromagnets, and spin–orbit coupled systems. 
We investigate spin-selective quantum transport in heterostructures composed of a normal metal and a two-dimensional $d$-wave altermagnet, and identify a universal mechanism for achieving perfect spin polarization. 
The mechanism is dictated by Fermi-surface geometry: closed surfaces in weak altermagnets yield partial and oscillatory spin filtering, whereas open surfaces in strong altermagnets intrinsically enforce fully spin-polarized conductance. 
Exploiting these distinct transport properties, we propose all-electrical spin-filter and spin-valve architectures, where resonant tunneling produces highly spin-polarized conductance tunable by gate voltage and interface transparency. 
Altermagnets with open Fermi surfaces further support gate-reversible perfect spin polarization that remains robust against interface scattering, disorder, and temperature. 
We also demonstrate an electrically controlled spin valve that reproduces the functionality of magnetic tunnel junctions without magnetic fields or relativistic mechanisms. 
$d$-wave altermagnets with open Fermi surfaces thus provide a new platform for low-dissipation, scalable, and magnetic-field-free spintronic devices, with potential for integration into next-generation quantum and CMOS-compatible technologies.
\end{abstract}

\maketitle
%%%%%%%%%%%%%%%%%%%%%%%%%%%%%%%%%%%%%%%%%%%%%%%%%%%%%%%%%%%%%%%%%%%%%%%%%%%%%
\section*{Introduction}
% revies of UM 
Unconventional magnets have emerged as a new class of systems that lie beyond the traditional dichotomy of ferromagnets and antiferromagnets \cite{Smejkal2022c,Shim2024a,Cheong2024,Jungwirth2024,Bai2024,Yan2024,Smejkal2022a,Smejkal2022,Hellenes2023,Jungwirth2024a,Yuri2025Superconducting}.
These so-called third type of magnets support spin-split Fermi surfaces resembling those of ferromagnets \cite{Hirohata2020Review,NatureMaterials2022New}, yet exhibiting globally vanishing magnetization due to a compensated magnetic ordering, akin to antiferromagnets \cite{DalDin2024Antiferromagnetic,Jungwirth2018Multiple,Baltz2016Antiferromagnetic}.
Depending on the momentum-space symmetry of the exchange field and the underlying crystalline symmetries, unconventional magnets can be classified according to the angular character of the spin splitting, including $p$-, $d$-, $f$-, $g$-, and $i$-wave types \cite{Smejkal2022a,Smejkal2022,Hellenes2023,Tagani2024,Ezawa2024b,LiuPRX2022SpinGroupSymmetry,ChenPRX2024EnumerationRepresentationTheory,ZhuNC2025Magneticgeometryinduced,ChenN2025Unconventionalmagnonscollinear,LiuNP2025Differentfacetsunconventional}.  
Among these, $d$-, $g$-, and $i$-wave magnets are known as altermagnets (AMs), characterized by simultaneously broken time-reversal symmetry and spatial rotational symmetry, but preserving their joint symmetry \cite{Smejkal2022a,Smejkal2022}. 
Altermagnets exhibit fully compensated spins and broken parity-time ($\bm{PT}$) symmetry. \cite{Cheong2024}.
Furthermore, AMs can also be classified according to their closed and open Fermi surfaces,
\cite{Smejkal2022a,Smejkal2022,Das2023Transport,Das2024Crossed,Nagae2025Spin}, leading to the concept of weak and strong AMs.
In contrast, $p$- and $f$-wave magnets break only rotational symmetry but preserve time-reversal symmetry originating from nonsymmorphic lattice symmetries \cite{Hellenes2023,Jungwirth2024a}, typically exhibiting closed anisotropic Fermi surfaces.
\begin{figure}
    \centering
    \includegraphics[width=1\linewidth]{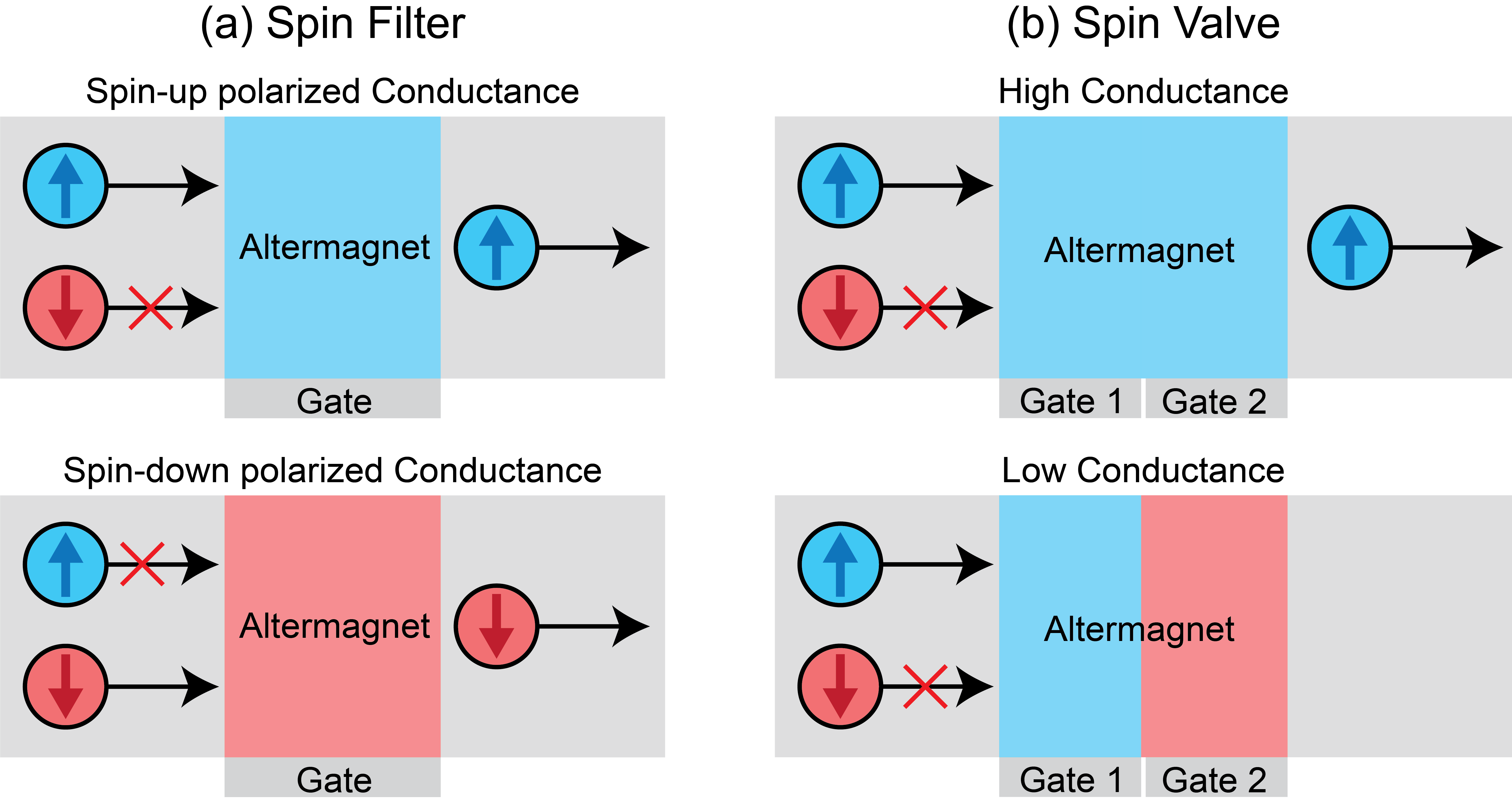}
    \caption{Schematic of electrically controlled spintronic devices based on AM heterostructures: (a) spin filter and (b) spin valve.
    (a) In the spin filter, a gate applied to the AM region generates fully spin-polarized conductance by selectively blocking one spin channel.  
    (b) The spin valve consists of two gated spin filters, exhibiting electrically tunable switching between high and low conductance states. 
    This behavior is analogous to the parallel and antiparallel magnetization configurations in conventional ferromagnetic bilayer spin valves, but achieved here without magnetic fields and net magnetization.
    }
    \label{fig0}
\end{figure}

Unconventional magnets are promising platforms for spintronic devices \cite{NatureMaterials2022New,DalDin2024Antiferromagnetic}.
The anisotropic spin split Fermi surface offers mechanisms for giant magnetoresistance \cite{ifmmode2022Giant}, spin-orbit-free anomalous Hall effects \cite{Feng2022,Smejkal2022c,GonzalezBetancourt2023,Dufouleur2023,Reichlova2024}, spin-transfer torques \cite{Bai2022Observation,Karube2022Observation,Han2024Harnessing}, spin filtering effects \cite{Samanta2024Spin}, spin pumping effects \cite{Sun2023Spin}, non-linear transports \cite{ZhuNC2025Magneticgeometryinduced}, light-matter interactions \cite{Werner2024High,Farajollahpour2025Light}, light-induced spin density \cite{Fu2025Floquet}, non-Hermitian electronic responses \cite{Reja2024,Dash2024}, strongly correlation in Mott insulators \cite{Maznichenko2024Fragile} and other novel phenomena \cite{Lin2025Coulomb,Chen2024Electrical}.
These mechanisms are expected in experimentally accessible altermagnetic materials exhibiting novel spin transport, such as RuO$_2$ \cite{Liu2024c,Feng2022,Dufouleur2023}, MnTe \cite{Hariki2024,GonzalezBetancourt2023,Reichlova2024,Rial2024}, Mn$_5$Si$_3$ \cite{Reichlova2024,Rial2024}, CrSb \cite{Ding2024,Yu2025Neel}, 
FeS$_2$ \cite{Li2025Altermagnetism}, MnF$_2$ \cite{Faure2025Altermagnetism}, Mn$_2$Au \cite{Elmers2020}, La$_{2}$O$_{3}$Mn$_{2}$Se$_{2}$, Ba$_{2}$CaOsO$_{6}$ \cite{Ubiergo2025Atomic} and La$_{2}$CuO$_{4}$ \cite{Brekke2023Two}. 

In addition to their normal-state properties, unconventional magnets have recently garnered attention in the context of superconducting spintronics by inducting high-parities spin-triplet states \cite{Fu2025Floquet,Maeda2025Classification,Fukaya2025Josephson,Chakraborty2024Constraints,Sukhachov2024Coexistence,Chatterjee2025Interplay}, orientation-dependent transport in altermagnetic-supercondctor hybridstructure \cite{Sun2023Andreev,Papaj2023Andreev, Nagae2025Spin,Zhao2025Orientation,Niu2024Orientation,Maeda2024Tunneling,Das2024Crossed,Nagae2025Spin,Niu2024Electrically}, gate-controlled Josephson junctions \cite{Fukaya2025Josephson,Zhang2024Finite,Lu2024Josephson,Sun2025Tunable,Ouassou2023dc,Beenakker2023Phase}, topological superconductors \cite{Zhu2023Topological,Chatterjee2025Interplay,Ghorashi2024Altermagnetic,Li2023Majorana,Li2024Realizing,Tanaka2024Theory}, nonreciprocal field-free superconducting devices \cite{Banerjee2024Altermagnetic,Cheng2024FieldFree,Chakraborty2024Perfect} and their superconducting phenomena generated by the interplay between unconventional magnetism and superconductivity \cite{Yuri2025Superconducting}.
Together, these developments establish unconventional magnets as a fertile ground for exploring spintronic functionalities without net magnetization.

At the moment, there exist several proposals for spintronic devices based on AMs \cite{Bai2024,Yan2024,Yuri2025Superconducting}.
For instance, spin-dependent tunneling barriers have been shown to induce spin splitting in insulating AMs \cite{Samanta2024Spin}.
Alternatively, $g$-wave altermagnetic semiconductors have demonstrated strain-induced spin-orbit coupling, enabling gate-controlled spin splitting \cite{Belashchenko2025Giant}. 
More recently, electrically gated spin-layer coupling can be achieved by applying different potentials to individual layers \cite{Zhang2024Predictable}. 
In recently developed ferroelectric-switchable AMs \cite{Duan2025Antiferroelectric,Gu2025Ferroelectric}, the spin-split Fermi surface can be reoriented by reversing the ferroelectric polarization. 
Despite these proposals offering a pathway toward AM-based spintronics, the unique non-relativistic nature of spin splitting in AM systems remains underutilized and largely unexplored.
Furthermore, most theoretical and experimental efforts have focused on weak AMs, while the transport properties of strong AMs, featuring open Fermi surfaces, remain largely underexplored \cite{Das2023Transport,Das2024Crossed,Nagae2025Spin}.
Thus, leveraging the intrinsic non-relativistic spin splitting of AMs for realizing spintronic functionalities without external magnetic fields remains an open and compelling challenge.

In this work, we explore spin-selective transport in heterostructures composed of AMs and normal metals, aiming to establish all-electrical spintronic functionality that harnesses the intrinsic, non-relativistic spin splitting of AMs.
Two AM-based spintronics devices with fundamental interest are proposed, including a spin filter and spin valve (Fig. \ref{fig0}).
In the regime of quantum-coherent transport, a conventional spin filter is realized in a ferromagnet junction, where the Zeeman splitting in combination with a build-up potential barrier gives rise to a spin-polarized flow since spin-up and spin-down electrons experience different barrier heights \cite{Streda2003Antisymmetric,Hirohata2020Review,NatureMaterials2022New}. 
While the spin polarization of the current through the junction is determined by the magnetization of the ferromagnet, the applied barrier can tune the allowed occupation of states belonging to the lower Zeeman level and thus strengthen the spin filtering.
In the proposed spin filter [Fig. \ref{fig0}(a)], a gate applied to the AM region generates switchable fully spin-polarized conductance by selectively blocking one spin channel, yet without requiring a Zeeman exchange field.  
Furthermore, two AM-based spin filters connected in series compose a spin valve [Fig. \ref{fig0}(b)].
This is similar to the conventional spin valve composed of a conventional magnetic bilayer, where only electrons with spin aligned to the magnetization contribute to the current \cite{NatureMaterials2022New,Hirohata2020Review}. 
Consequently, conductance is high when the magnetization directions of the two ferromagnetic layers are aligned, while it is suppressed when these directions are opposite, known as the parallel and antiparallel configurations, respectively.
Parallel-antiparallel switching is typically achieved by applying an external magnetic field, which controls the relative magnetization orientations between layers.
The resulting difference in conductance and thus the resistance under field reversal gives rise to the giant magnetoresistance effect \cite{Baibich1988GMR,Binasch1989GMR}, which forms the basis for modern electronics applications such as data storage technologies, magnetic random access memory, and magnetic field sensors \cite{Hirohata2020Review,NatureMaterials2022New,DalDin2024Antiferromagnetic,Jungwirth2018Multiple,Baltz2016Antiferromagnetic}.
The behaviors analogous to the parallel-antiparallel switching are also found in the proposed spin valve based on strong AMs, where the alteration of the spin polarization in each spin filter is achieved by the applied gate [Fig. \ref{fig0}(b)], without an external magnetic field and net magnetization. 

To elaborate on the mechanism of these devices, we begin by classifying weak and strong AMs based on the competition between the spin-dependent exchange interaction and kinetic energy (Fig. \ref{fig1}). 
We find that their distinct Fermi surfaces give rise to qualitatively different transport characteristics when assembled into junctions (Fig. \ref{fig2}). 
In particular, quantum resonant tunneling leads to spin-selective conductance that can be controlled by a gate voltage and interface transparency. 
For weak AMs, spin filtering arises through barrier-enhanced conductance polarization due to the suppression of one spin species by the confinement effect \cite{Desai2010Quantum,Cayao2015SNS,Cayao2021Confinement,Rainis2014Conductance}. 
In contrast, strong AMs enable fully spin-polarized conductance over wide energy ranges due to the intrinsic absence of one spin species near the Fermi level. 
We find that this intrinsic spin-filtering effect is robust against interface scattering and does not require magnetic fields or spin-orbit coupling (Fig. \ref{fig3}).
Based on this mechanism, we further propose a double-gated spin valve device in a strong AM (Fig. \ref{fig4}).
Interestingly, in this device, alternating 
between high and low conductance states can be realized by electrical gating, analogous to parallel and antiparallel spin configurations in conventional magnetic spin valves \cite{NatureMaterials2022New,DalDin2024Antiferromagnetic} but without requiring net magnetization or an external magnetic field.
Our results reveal that the non-relativistic spin splitting of AM offers a promising ground for engineering robust and tunable spintronic devices based entirely on electric field control. 
The proposed all-gate-controlled altermagnetic spin filter and spin valve are enabled by the intrinsic Fermi surface structure of weak and strong altermagnetic materials, without the need for net magnetization, relativistic spin-orbit coupling, or external magnetic fields. 
% These mechanisms can be implemented in experimentally accessible altermagnetic materials such as RuO$_2$ \cite{Liu2024c,Feng2022,Dufouleur2023}, MnTe \cite{Hariki2024,GonzalezBetancourt2023}, Mn$_5$Si$_3$ \cite{Reichlova2024,Rial2024}, CrSb \cite{Ding2024}, and Mn$_2$Au \cite{Elmers2020}, offering promising platforms for altermagnet-based spintronics.

% The remainder of this work is organized as follows. 
% The band structure, Fermi surface features of weak and strong AM are demonstrated in Sec. \ref{sec2}, which determines the transport properties and the spin filter effect in the altermagnetic junction illustrated in Sec. \ref{sec3}.
% Based on these properties, two spintronic devices, gate-controlled spin filter and spin valve, are proposed in Sec. \ref{sec4}. 
% The conclusion is given in Sec. \ref{sec5}.

\section*{Results}
\subsection*{Weak and strong altermagnets} \label{sec2}
We start by introducing the Hamiltonian of $d$-wave AMs in continuum limit in the spin basis $\left( \psi_{\bm{k},\uparrow}, \psi_{\bm{k},\downarrow} \right)^{\mathrm{T}}$ \cite{Smejkal2022a, Smejkal2022}
\begin{equation}
H_{\text{AM}}( \bm{k}) =\left( 
\begin{array}{cc}
H_{+}( \bm{k})  & 0 \\ 
0 & H_{-}( \bm{k}) 
\end{array}%
\right) \text{,}  \label{eq_hsigma1}
\end{equation}%
where
\begin{equation}
H_{\sigma}(\bm{k}) = \xi_{\bm{k}} + J_{\bm{k}}, \label{eq_hsigma}
\end{equation}
is the spin-resolved Hamiltonian for electrons with spin $\sigma = +1$ ($\sigma = -1$) being parallel (antiparallel) to the N\'eel vector, which is hereafter considered along $z$. 
The kinetic energy $\xi_{\bm{k}}$ and altermagnetic exchange field $J_{\bm{k}}$ are defined, respectively, as
\begin{eqnarray}
\xi_{\bm{k}} &=&ta^{2}\left( k_{x}^{2}+k_{y}^{2}\right) \text{,}
\label{eq_xik} \\
J_{\bm{k}} &=&\sigma Ja^{2}\left[ \left( k_{x}^{2}-k_{y}^{2}\right) \cos
2\theta_{J}+2k_{x}k_{y}\sin 2\theta_{J}\right] \text{.} \label{eq_jk}
\end{eqnarray}
Here, $t $ is the nearest-neighbor hopping energy (set as the energy unit $t=1$), and $a$ is the lattice constant (set as $a=1$). 
The two-dimensional momentum is denoted by $\bm{k} = (k_{x}, k_{y})$, with magnitude $k = |\bm{k}|$ and polar angle $\phi_{k} = \arctan(k_{y}/k_{x})$. 
The parameter $J$ represents the strength of the altermagnetic field, and $\theta_{J}$ defines the angle between the altermagnetic lobe and the $x$-axis. 

The momentum-dependent spin splitting is captured by Eq. (\ref{eq_jk}), which describes a $d_{x^{2}-y^{2}}$-wave ($d_{xy}$-wave) AM when $\theta_{J}=0$ ($\theta_{J}=\pi /4$)~\cite{Smejkal2022a,Smejkal2022}, hosting an anisotropic spin split $d$-wave Fermi surface that depends on the momentum direction $\phi_{k}$ for a given $\theta_{J}$.  
The maximum altermagnetic effect occurs along momentum directions $\phi_{k} = \theta_{J} + n\pi/2$, with $n \in \mathbb{Z}$, whereas it vanishes along directions $\phi_{k} = \theta_{J} + (2n + 1)\pi/4$.

\begin{figure}
    \centering
    \includegraphics[width=1\linewidth]{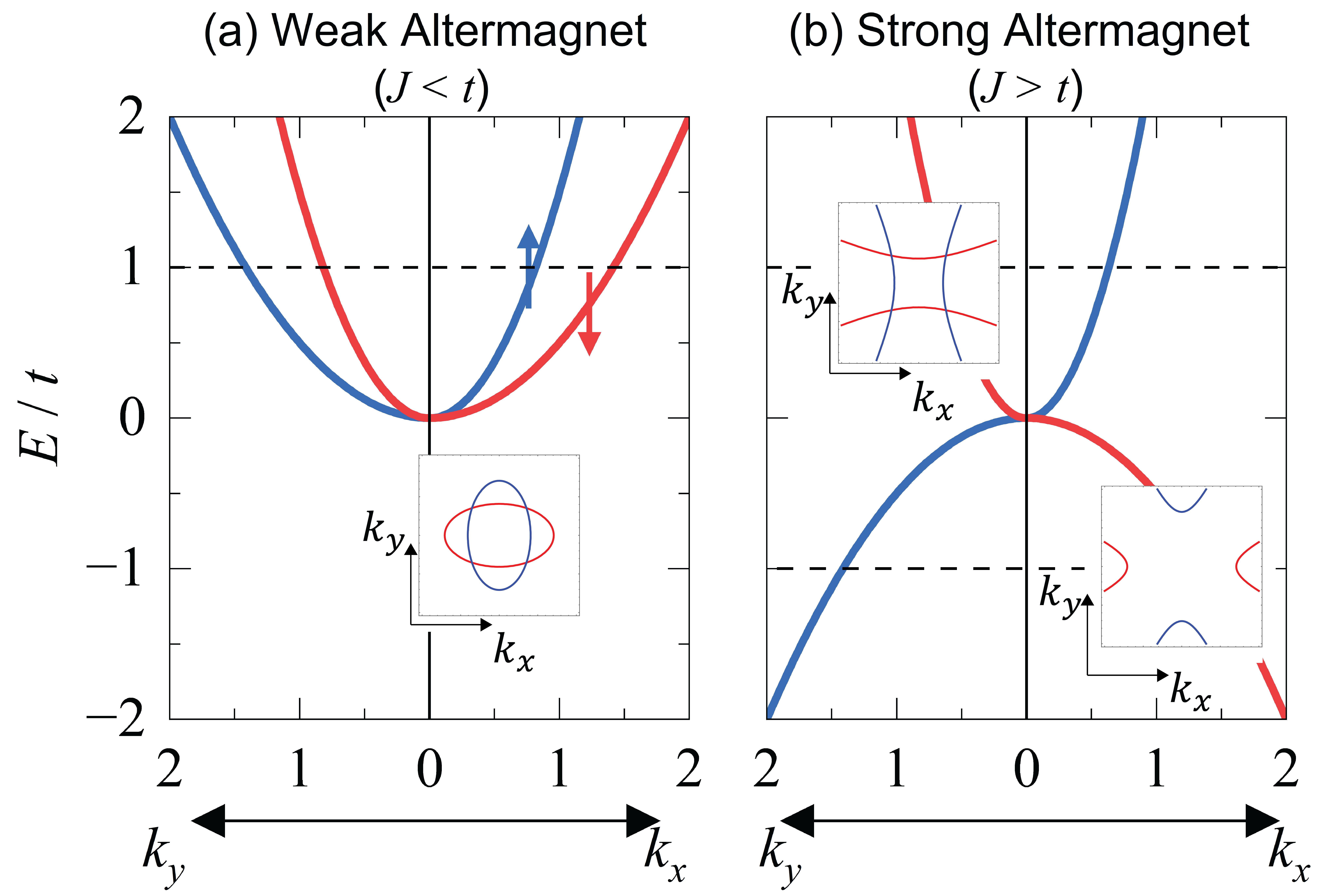}
    \caption{Dispersion of (a) weak and (b) strong AM along $k_x$ and $k_y$ directions with $k_y=0$ and $k_x=0$, respectively. 
    The insets exhibited the anisotropic Fermi surface at $E/t=1$ for weak altermagnet and $E/t=\pm1$ for strong altermagnet as indicated by the horizontal dashed line.
    The spin-up (spin-down) subband is in blue (red).
    The altermagnetic strength is (a) $J/t=0.5$ and (b) $J/t=1.5$.
    Parameters: $t=1$ is the energy unit and $a=1$ is the lattice constant.}
    \label{fig1}
\end{figure}

The competition between the altermagnetic term, Eq. (\ref{eq_jk}), and the kinetic energy term, Eq. (\ref{eq_xik}), defines weak and strong altermagnetism. 
To illustrate this effect, we first consider the $d_{x^{2}-y^{2}}$-wave AM as an example, whose dispersion can be obtained from Eq.(\ref{eq_hsigma}) with $\theta_{J}=0$, as
\begin{equation}
E_{\sigma }=\left( t+\sigma J\right) a^{2}k_{x}^{2}+\left( t-\sigma J\right)
a^{2}k_{y}^{2}\text{.} \label{eq_dis}
\end{equation}
Figure \ref{fig1} displays $E_{\sigma }$ [Eq. (\ref{eq_dis})] as a function of $k_{x,y}$.
There are two directions of maximal altermagnetic effect, namely along the $k_x$ and $k_y$ axes for $k_y = 0$ and $k_x = 0$, respectively. 
Along these directions, the dispersion of a weak AM ($J < t$) \cite{Das2024Crossed} consists of two upward-opening parabolas with spin- and direction-dependent curvature [Fig. \ref{fig1}(a)]. 
For a given energy $E$, the Fermi surface satisfies the equation of a standard ellipse,
\begin{equation}
\frac{a^{2}k_{x}^{2}}{E/\left( t+\sigma J\right) }+\frac{a^{2}k_{y}^{2}}{%
E/\left( t-\sigma J\right) }=1\text{,}
\end{equation}
which describes two spin-dependent elliptical Fermi surfaces with mutually orthogonal principal axes, with semi-major and semi-minor axes being $\sqrt{E/(t+\sigma J)}/a$ and $\sqrt{E/(t-\sigma J)}/a$ depending on spin [see the inset of Fig. \ref{fig1}(a)].

% As shown in the inset of Fig. \ref{fig1}, the maximum altermagnetic effect occurs along momentum directions $\phi_{k} = \theta_{J} + n\pi/2$, with $n \in \mathbb{Z}$, whereas it vanishes along directions $\phi_{k} = \theta_{J} + (2n + 1)\pi/4$.

In contrast, for a strong AM ($J > t$) \cite{Das2024Crossed}, the dispersion consists of two parabolic bands with opposite opening directions and spin-dependent curvature [see Fig. \ref{fig1}(b)]. 
This arises because the sign of the effective mass, determined by the coefficients $(t \pm \sigma J)a^2$ in Eq. (\ref{eq_dis}), changes along orthogonal momentum directions. 
As a result, the band structure hosts a saddle point \cite{Ziletti2015Van} at the spin-degenerate momentum $(k_x, k_y) = (0, 0)$, where the curvature is positive in one direction and negative in the perpendicular one. 
Such a saddle point gives rise to a logarithmic van Hove singularity in the density of states near the corresponding energy \cite{Ziletti2015Van}.
The opposite opening directions of the spin-dependent bands allow electronic states to exist below $E = 0$, with spin-up and spin-down states shifted oppositely in energy [Fig. \ref{fig1}(b)]. 
This energy asymmetry arises from the spin-dependent curvature of the bands in orthogonal directions. 
As a result, the strong AM is characterized by open, spin-dependent anisotropic Fermi surfaces, which take the form of two orthogonally oriented sets of hyperbolas. 
These Fermi contours, distinct from the closed elliptical shapes found in the weak AM case, are described by
\begin{equation}
\sigma \frac{a^{2}k_{x}^{2}}{E/\left( \sigma t+J\right) }-\sigma \frac{%
a^{2}k_{y}^{2}}{E/\left( J-\sigma t\right) }=1\text{,}
\end{equation}
for a fixed energy $E$. 
These open Fermi surfaces impose directional constraints on the availability of spin-polarized electronic states. 
As shown in the inset of Fig. \ref{fig1}(b), for a given positive energy $E$, spin-up (spin-down) states are absent when $a|k_x| < \sqrt{E/(t + J)}$ [$a|k_y| < \sqrt{E/(t + J)}$], due to the absence of real solutions for the corresponding hyperbolic Fermi contour. 
Conversely, for negative energies $E < 0$, spin-up (spin-down) states are absent when $a|k_y| < \sqrt{|E|/(t - J)}$ [$a|k_x| < \sqrt{|E|/(t - J)}$]. 
This directional open Fermi surface is a hallmark of the strong altermagnetic regime and stands in sharp contrast to the weak AM case, where both spin species coexist over closed elliptical Fermi surfaces. 
While our discussions above are based on $d_{x^{2}-y^{2}}$-wave AM with $\theta_J=0$, the distinct properties such as the dispersions, Fermi surface geometry between weak and strong AM, are shared in all $d$-wave AM with arbitrary $\theta_J$.
% As we discuss below, this fundamental distinction leads to different spin transport characteristics between weak and strong AM junctions \cite{Das2024Crossed,Nagae2025Spin,Das2023Transport}.
Remarkably, the distinct Fermi-surface geometries described by the continuum altermagnetic Hamiltonian [Eq.\,\ref{eq_hsigma1}] also emerge in lattice-model Hamiltonian fitted to real materials such as RuO$_2$ \cite{Liu2024c,Feng2022,Dufouleur2023}, La$_2$O$_3$Mn$_2$Se$_2$, and Ba$_2$CaOsO$_6$ \cite{Ubiergo2025Atomic} (see Sec.\,I of the Supplemental Material). 
In the following, we therefore restrict our investigation to the continuum model, which clearly illustrates the underlying mechanism of spin-selective transport. 
This mechanism is further corroborated by the lattice model results presented in the Discussion section.

\subsection*{Spin-Selective Transport in Altermagnetic Junctions} \label{sec3}

\begin{figure*}
    \centering
    \includegraphics[width=0.99\linewidth]{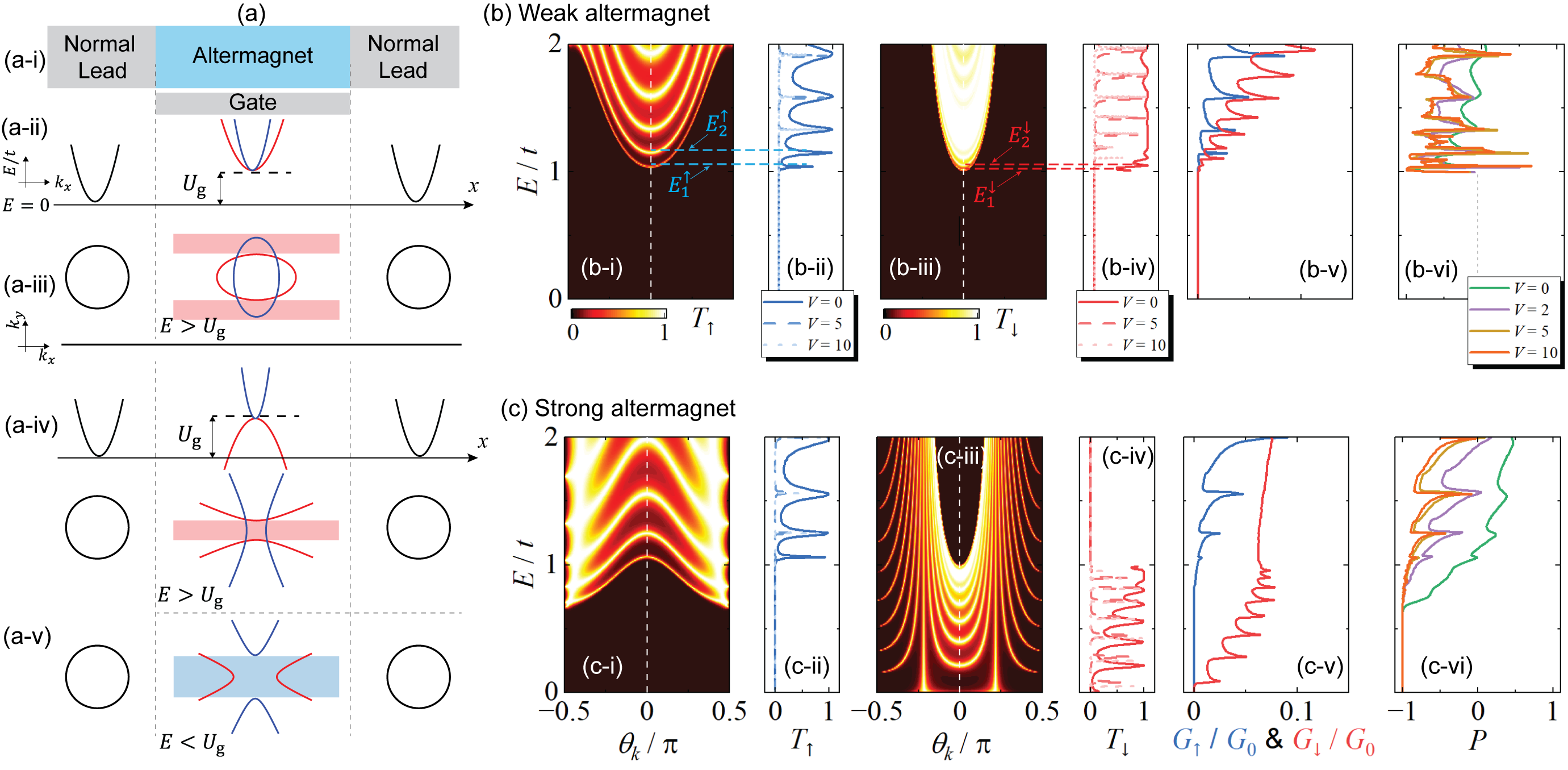}
    \caption{
    (a) Schematic illustration of the spin-filter effect.
    (a-i) The spin-filter is composed of normal metals and a gated altermagnet by $U_g$. 
    The junction is along the $x$-direction, keeping periodic boundary conditions along the $y$-direction.
    (a-ii) The normal-metal leads exhibit spin-degenerate parabolic dispersions (black lines), while the weak AM features spin-dependent parabolic bands with the same curvature direction, with blue and red lines denoting the spin-up and spin-down states, respectively.
    The dispersions are exhibited along $k_x$ direction with $k_y=0$. 
    (a-iii): Isotropic Fermi surface of normal-metal leads and anisotropic closed and open spin-resolved Fermi surfaces for weak AMs.
    The Fermi surfaces are exhibited in $k_x$-$k_y$ plane.
    Blue (red) shaded regions indicate transverse mode ($k_y$) ranges forbidden for spin-up (spin-down) electrons due to the AM band structure.    
    (a-iv) and (a-v) are the same as (a-ii) and (a-iii), respectively, but for strong AMs. 
    (b) Transmission probability, spin-resolved conductance and spin polarization for a weak AM. 
    (b-i) and (b-iii): Energy and incident-angle ($\theta_k$) resolved transmission probabilities $T_{\uparrow}$ and $T_{\downarrow}$. 
    (b-ii) and (b-iv): Transmission for normal incidence ($\theta_k = 0$) at various tunneling barrier heights $V$.
    The horizontal dashed line denotes the resonant energy levels [Eq. (\ref{eq_Ens})] due to the confinement condition [Eq. \ref{eq_resonant}].
    (b-v) and (b-vi): Energy-dependent spin-resolved conductance and resulting spin polarization. 
    (c) Same as panel (b), but for a strong AM. 
    Parameters: $U_L = 0$, $U_g = 1$, $V = 0$ in (i, iii, v), and junction length $d = 20a$. 
    The $d_{x^2-y^2}$-wave AM with $\theta_J=0$ is considered.
    }
    \label{fig2}
\end{figure*}

The contrasting spin-split Fermi surfaces in weak and strong AMs give rise to qualitatively distinct transport responses for designing spintronic devices. 
A central objective in spintronics is generating and manipulating spin-polarized conductance or current, particularly through electrically tunable spin filters \cite{Hirohata2020Review,NatureMaterials2022New,DalDin2024Antiferromagnetic,Jungwirth2018Multiple,Baltz2016Antiferromagnetic}. 
To explore this device, we consider a junction composed of an AM attached with two normal metal leads, as schematically shown in Fig. \ref{fig2}(a-i). 
This heterostructure supports a gate-controllable spin-filtering effect, enabled by the intrinsic energy- and momentum-dependent spin splitting in the altermagnetism.
Using a quantum scattering formalism (see Methods), the spin-resolved transmission probability is given by
\begin{equation}
T_\sigma(E, \theta_k) = \left| \tau_\sigma(E, \theta_k) \right|^2,
\label{eq_Ts}
\end{equation}
and the spin-resolved conductance at zero temperature reads \cite{Datta1997Electronic}
\begin{equation}
G_\sigma(E_F) = G_0 \int_{-\pi/2}^{\pi/2} T_\sigma(E, \theta_k)\, \cos \theta_k \, d\theta_k,
\label{eq_Gs}
\end{equation}
where $G_0 = e^2 W k_F / (2\pi h)$ is the conductance quantum per spin, $W$ denotes the sample width, $\tau_\sigma$ is the transmission amplitude given by Eq. (\ref{eq_taus}). 
The total conductance of the junction is given by the sum over spin channels:
\begin{equation}
G = G_\uparrow + G_\downarrow,
\end{equation}
where the spin indices $\uparrow$ and $\downarrow$ correspond to $\sigma = +1$ and $\sigma = -1$, respectively. 
In the linear response regime, the spin-resolved current under an applied bias $V_{\text{bias}}$ is $I_{\sigma} = G_{\sigma} V_{\text{bias}}$, and the total current follows as $I = I_\uparrow + I_\downarrow$ \cite{Datta1997Electronic}.
To quantify the relative contribution of each spin channel, we define the spin polarization as
\begin{equation}
P = \frac{G_\uparrow - G_\downarrow}{G_\uparrow + G_\downarrow},
\label{eq_P}
\end{equation}
where $P > 0$ ($P < 0$) indicates a dominant contribution from spin-up (spin-down) electrons.
The spin polarization $P$ serves as a measure of spin-filtering efficiency when a spin-degenerate electron beam is injected. 
In particular, $P = +1$ ($P = -1$) corresponds to a fully spin-polarized conductance and current carried entirely by spin-up (spin-down) states. 
The generation and electrical control of spin-polarized conductance and thus the current are essential goals in the development of spintronic devices.

\subsubsection*{Transport in weak altermagnetic junctions} \label{sec3b}

Based on the quantum scattering formalism, in the following, we analyze transport through a spin filter based on a weak $d_{x^2-y^2}$ [Fig. \ref{fig2}(a-ii) and (a-iii)], characterized by a closed, spin-dependent Fermi surface [Fig. \ref{fig1}(a)]. 
Figure \ref{fig2}(b) presents the energy- and angle-resolved transmission probabilities, as defined in Eq. (\ref{eq_Ts}), along with the resulting spin-resolved conductance [Eq. (\ref{eq_Gs})] and spin polarization [Eq. (\ref{eq_P})].

The spin-dependent transmission probabilities $T_{\uparrow}$ and $T_{\downarrow}$ are shown in Figs. \ref{fig2}(b-i) and \ref{fig2}(b-iii), respectively, as functions of the incident energy $E$ and injection angle $\theta_k$. 
The corresponding normal-incidence transmission profiles, $T_{\uparrow}(\theta_k = 0)$ and $T_{\downarrow}(\theta_k = 0)$, are plotted in Figs. \ref{fig2}(b-ii) and \ref{fig2}(b-iv) for different values of the tunneling barrier height $V$.
Resonant transmission ($T_{\sigma} = 1$) occurs when the junction length $d$ satisfies the standing-wave condition \cite{Desai2010Quantum}, such that it equals an integer multiple of half the wavelength of spin-$\sigma$ electrons in the altermagnetic region (See Method). 
The corresponding wavelength is given by $\lambda_{\sigma} = 2\pi / k_{+}^{\sigma}$, where $k_{+}^{\sigma}$ [see Eq.~(\ref{eq_kspm})] denotes the propagating wave vector of spin-$\sigma$ electrons in the altermagnetic layer.
This results in the resonance condition,
\begin{equation}
k_{+}^{\sigma} d = n\pi, \quad (n = 1, 2, 3, \cdots), \label{eq_resonant}
\end{equation}
which ensures constructive interference between the left-going and right-going 
propagating modes in AM in Fabry–P\'erot–like structures \cite{Desai2010Quantum,Cayao2015SNS,Cayao2021Confinement,Rainis2014Conductance}. 
So that the phase difference accumulated over distance $d$ between the two modes is an integer multiple of $\pi$.
With this condition, one can check that the denominator of Eq. (\ref{eq_taus}) becomes minimal, resulting in maximal transmission probability.
% By substituting Eq. (\ref{eq_kspm}) into the resonance condition [Eq. (\ref{eq_resonant})], 
The corresponding resonant energy levels related to the confined standing wave are 
\begin{equation}
E_{n}^{\sigma}(\theta_k) = t \frac{t_+ (n\pi a / d)^2 + U_g + t_- \sin^2 \theta_k \, U_L / t}{t - t_- \sin^2 \theta_k}, \label{eq_Ens}
\end{equation}
with $t_\pm = t \pm \sigma J$.
Here, $U_L$ sets the chemical potential in the left ($x < 0$) and right ($x > d$) normal leads, $U_g$ denotes the gate voltage applied to the central AM region, and $V$ represents the tunneling barriers at the normal–AM interfaces located at $x = 0$ and $x = d$. 
The resonant energy [Eq.\,(\ref{eq_Ens})] levels depend on the spin index $\sigma$, the transverse incident angle $\theta_k$, and device parameters such as $U_L$, $U_g$, and $d$, but are independent of the tunneling barrier $V$. 
Thus, the transmission probability remains unity even in the presence of a finite tunneling barrier $V$ when the incident energy satisfies the resonance condition, $E = E_n^{\sigma}(\theta_k)$, but becomes strongly suppressed in the off-resonant regime, $E \in (E_n^{\sigma}, E_{n+1}^{\sigma})$ [See Fig. \ref{fig2}(b-i)-(b-iii) for $n=1$].
The energy spacing between neighboring resonant levels is
\begin{eqnarray}
\delta_{n,E}^{\sigma}(\theta_k) &=& E_{n+1}^{\sigma}(\theta_k) - E_{n}^{\sigma}(\theta_k) \label{eq_dE} \\
&=& t a^2 \pi^2 \frac{t + \sigma J}{t - (t - \sigma J) \sin^2 \theta_k} \frac{2n + 1}{d^2}, \nonumber
\end{eqnarray}
which is spin-dependent. 
The level spacing decreases with increasing junction length $d$, but increases with the incident angle $\theta_k$. 
Since $\delta_{E}^{+}(\theta_k) > \delta_{E}^{-}(\theta_k)$, finite $T_{\downarrow}$ may appear in energy regions where $T_{\uparrow}$ is suppressed [See Fig. \ref{fig2}(b-i)-(b-iii)].  
This asymmetry spin channel leads to spin-polarized tunneling.
The degree of suppression increases with $V$, as the barrier height reduces the amplitude of evanescent modes within the altermagnetic region and enhances the energy selectivity of the resonance, thereby narrowing the transmission windows and increasing the contrast between on- and off-resonant transport.
This behavior is illustrated in Fig. \ref{fig2}(b-ii) and Fig. \ref{fig2}(b-iv), which show the normal-incidence transmission probabilities $T_{\uparrow}(\theta_k = 0)$ and $T_{\downarrow}(\theta_k = 0)$, respectively, for various values of the tunneling barrier $V$. 
As $V$ increases, transmission away from resonance is progressively suppressed, while the resonant peaks remain robust, highlighting the role of $V$ in enhancing spin-dependent energy filtering.
In the normal incident case, the lowest resonant peak occurs at $E = E_1^{\sigma}(0) = (t + \sigma J)(\pi a/d)^2 + U_g \approx U_g$ for long junctions ($d \gg a$) and tunneling for $E < U_g$ is negligible since the incident electron energy lies below the bottom of the spin-dependent parabolic dispersion set by $U_g$ [Fig. \ref{fig1}(a-ii)]. 
For general injection angles, Eq. (\ref{eq_Ens}) implies that only electrons within a restricted angular range contribute to the transmission. 
This angular window is bounded by a critical angle $\theta_c^{\sigma}$ for spin $\sigma$, given by
\begin{equation}
\theta_c^{\sigma} = \arcsin \sqrt{ \frac{E - U_g}{\left(1 - \sigma J/t\right)(E + U_L)} }.
\end{equation}
Only electrons with incident angle $\theta_k^{\sigma} < \theta_c^{\sigma}$ can contribute to transmission. 
This spin-dependent critical angle arises from the anisotropic spin-split Fermi surface of AMs. 
For instance, in a $d_{x^2 - y^2}$ AM with $\theta_J = 0$, and a junction oriented along the $x$-direction [Fig. \ref{fig2}(a-iii)], spin-up electrons exhibit a broader angular transmission window compared to spin-down electrons.
 
By summing over all possible incident directions, the spin-resolved conductance [Fig. \ref{fig2}(b-v)] exhibits the spin-dependent resonance condition described by Eq. (\ref{eq_resonant}).
Specifically, the spin-up conductance is significantly suppressed by the tunneling barrier except for the resonant peaks, while the spin-down conductance remains measurable. 
These resonant features correspond to constructive interference conditions for the spin-dependent wavefunctions inside the altermagnetic region.
% , where the electrons 
% This behavior originates from the spin-dependent effective mass in weak AMs, which arises from the anisotropic $d$-wave spin splitting. 
% The resulting energy dispersion supports spin-selective tunneling due to the mismatch in propagation momentum, $k^\sigma_{+}$, between spin-up and spin-down carriers. 
As the tunneling barrier height $V$ increases, the off-resonant transmission is further suppressed, thereby enhancing the spin polarization $P$ [Fig. \ref{fig2}(b-vi)].
However, because of the continuous and overlapping band structure of the weak AM, both spin channels inevitably contribute to transport for a given energy, see Fig. \ref{fig2}(b-i) and (b-iii). 
Electrons incident at oblique angles ($\theta_k \neq 0$) retain finite tunneling probabilities. 
This leads to a residual spin-up contribution that limits the maximum achievable spin polarization.
Consequently, even in the presence of a high tunneling barrier, full spin polarization ($P = 1$ or $-1$) is hardly achieved. 
Instead, the spin polarization saturates at approximately $P \sim 80\%$ for large $V$, as demonstrated in Fig. \ref{fig2}(b-vi). 

The partial polarization in weak AM discussed above poses two notable challenges for device applications: 
(i) The maximum spin polarization is not fully tunable and remains sensitive to the Fermi energy $E$. 
This limits the flexibility of weak AM-based spin filters compared to conventional ferromagnetic systems \cite{Hirohata2020Review,NatureMaterials2022New}, where full spin polarization can be realized and externally modulated through magnetic fields \cite{Moodera1995Large} or spin-transfer torques \cite{Ralph2008Spin}.
(ii) Enhancing spin filtering requires a high tunneling barrier, which inherently reduces the total conductance and the current. 
The inverse relationship between spin polarization and conductance imposes a fundamental constraint on device performance, especially in regimes where a sizable spin-polarized current is required for detection or functionality.
While the challenge (i) can be mitigated by applying gate voltages to shift the resonance conditions as demonstrated in Sec. \ref{sec4a}, the challenge (ii) is intrinsic to the weak AM band structure. 
However, as we show in the following, this limitation can be circumvented by employing strong AM materials. 
In that case, the emergence of spin-polarized transport arises from the intrinsic band structure of the strong AM itself, rather than relying on interfacial tunneling effects. 
This results in robust and fully spin-polarized transport, even in the absence of a tunneling barrier.

\subsubsection*{Transport properties in strong altermagnetic Junctions} \label{sec3c}

The distinct dispersion of a strong AM [Fig. \ref{fig1}(b)] enables a fundamentally different electrically controlled transport behavior compared to its weak counterpart. 
In this case, we note that the resonant conditions in Eqs. (\ref{eq_resonant})–(\ref{eq_dE}) still apply.
In particular, the spin polarization of the Fermi surface in a strong AM can be switched by tuning the energy, as illustrated in Fig. \ref{fig2}(a-iv), which is a behavior absent in weak AMs.
As a result, spin-down (spin-up) electrons with $\theta_k^- < \theta_c^-$ ($\theta_k^+ < \theta_c^+$) are blocked for $E > U_g$ ($E < U_g$), as shown in Fig. \ref{fig2}(c-ii) and Fig. \ref{fig2}(c-iii), respectively. 
This can also be understood from the possible Fermi surface contours at positive energy $E$, illustrated in the inset of Fig. \ref{fig1}(b). 
In this regime, spin-up (spin-down) states are absent when $a|k_x| < \sqrt{(E - U_g)/(t + J)}$ ($a|k_y| < \sqrt{(E - U_g)/(t + J)}$). 
Conversely, at negative energies, spin-up (spin-down) states vanish when $a|k_y| < \sqrt{(E - U_g)/(t - J)}$ ($a|k_x| < \sqrt{(E - U_g)/(t - J)}$), see Fig. \ref{fig2}(a-iv). 

Consequently, transmission at $E > U_g$ ($E < U_g$) is dominated by spin-up (spin-down) electrons, as shown in Fig. \ref{fig2}(c-i)-(c-iv). 
This feature is most pronounced for normally incident electrons, which is exclusively carried by spin-up (spin-down) states for $E > U_g$ ($E < U_g$).
Spin-up electrons with $\theta_k \neq 0$ are blocked for $E < U_g$ due to the large forbidden angle range, leading to a fully spin-polarized transmission and thus conductance [Eq. (\ref{eq_Gs})] in this energy regime, while for $E > U_g$, spin-down electrons with $\theta_k > \theta_c^-$ can still contribute to transmission at $E > U_g$, thereby reducing the degree of spin polarization.  
This behavior is reflected in the conductance and spin polarization shown in Fig. \ref{fig2}(c-v) and Fig. \ref{fig2}(c-vi).

\subsubsection*{Effects of the altermagnetic orientation $\theta_J$}

We now examine the influence of the altermagnetic orientation angle $\theta_J$ on the resulting spin polarization. 
For a $d_{x^2 - y^2}$-wave weak AM with $\theta_J = 0$, our analysis reveals that spin-polarized transport favors the spin-down channel, yielding $P < 0$, tunable via the incident electron energy $E$ and the barrier height $V$ [see Fig. \ref{fig2}(b-vi)]. 
Rotating the orientation to $\theta_J = \pi/2$, corresponding to an equivalent $d_{x^2 - y^2}$-wave AM rotated by $90^\circ$, reverses the spin polarization, resulting in $P > 0$. 
This configuration can be interpreted as reorienting the AM junction along the $y$-direction rather than the $x$-direction in Fig. \ref{fig2}.
The reversal spin polarization arises from the interchange of spin-split subbands, effectively mapping the spin index $\sigma \rightarrow -\sigma$ in key transport signatures such as resonant energy levels and critical incident angles, thereby inverting the sign of $P$. 
In contrast, for a $d_{xy}$-wave AM with $\theta_J = \pi/4$, transport is dominated by spin-degenerate propagation along the nodal directions ($y = \pm x$), resulting in spin-independent transmission and vanishing net spin polarization.

Orientation-dependent transport reveals their alternating behavior of AM between ferromagnetic and antiferromagnetic characteristics. 
Transport along directions of maximal spin splitting (e.g., $\theta_J = 0$ or $\pi/2$) exhibits ferromagnetic-like features, whereas spin-degenerate directions (e.g., $\theta_J = \pi/4$) give rise to antiferromagnetic-like effects. 
This anisotropy leads to orientation-dependent suppression of Andreev reflection \cite{Sun2023Andreev,Papaj2023Andreev,Niu2024Orientation,Zhao2025Orientation}, the emergence of $\pi$-state Josephson currents, and nonreciprocal diode effects \cite{Lu2024Josephson,Sun2025Tunable,Ouassou2023dc,Beenakker2023Phase}, all without the need for an external magnetic field.
The intrinsic spin splitting in $d_{x^2 - y^2}$-wave AM thus provides a highly tunable spin-selective transport, crucial for realizing spintronic functionalities in systems with zero net magnetization.

\subsection*{Gate-Controlled Spin Filtering and Spin Valve Effect in Altermagnet junctions} \label{sec4}

The distinct energy dependence of the spin-resolved transmission and the resulting conductance (Fig. \ref{fig2}), enables gate-tunable spin transport in AM-based junctions, offering a promising platform for spintronic applications without net magnetization. 
In this section, we propose two electrically controlled spintronic devices that utilize the anisotropic spin splitting in $d_{x^2 - y^2}$-wave AMs: a spin filter [Fig. \ref{fig2}(a) and Fig. \ref{fig3}] and a spin valve [Fig. \ref{fig4}]. 
In both cases, the spin-resolved conductance $G_\sigma$ and spin polarization $P$ can be modulated by gate voltages applied within the AM region and at the tunneling barrier at the AM–lead interface.
Notably, by exploiting the opposite band curvature of the spin-split subbands in strong AMs, we demonstrate a double-gate-controlled spin valve. 
This device exhibits a lower (higher) conductance state when the two gates are opposite (aligned), analogous to conventional spin valves based on ferromagnetic bilayers \cite{Hirohata2020Review,NatureMaterials2022New}, but realized here in a system with zero net magnetization.
% This configuration is similar to the parallel and anti-parallel structure in the conventional spin valve composed of two ferromagnetic layers separated by a non-magnetic or semiconducting spacer, where the relative orientations of the magnetization directions of adjacent ferromagnetic layers are controlled through the application of an external magnetic field. 

\subsubsection*{Gate-controlled spin filter effects} \label{sec4a}

\begin{figure}
    \centering
    \includegraphics[width=1\linewidth]{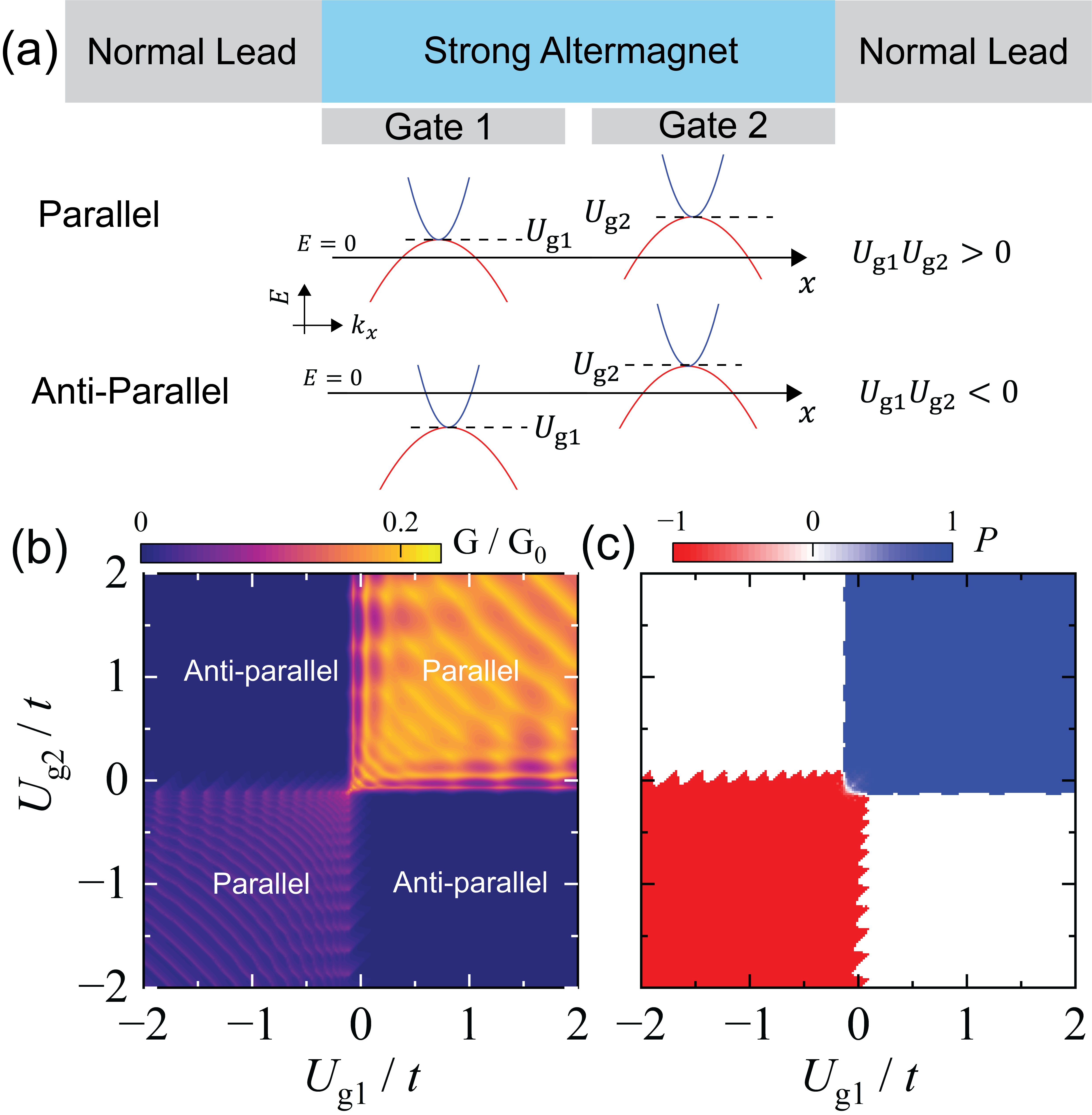}
 \caption{
Gate-controlled spin-filter effect for (upper panel) weak and (lower panel) strong altermagnets, corresponding to Fig. \ref{fig2}. 
(a) and (c): Spin-resolved conductance as a function of gate voltage $U_g$ at fixed tunneling barrier $V = 5$. 
(b) and (d): Spin polarization as a function of $U_g$ for various values of $V$. 
The Fermi level is set to $E = 0.1t$, and all other parameters are identical to those in Fig. \ref{fig2}.
}
    \label{fig3}
\end{figure}

The distinct energy-dependent conductance and polarization of strong and weak AMs enable gate- and barrier-controlled generation of spin-polarized conductance. 
To show this effect, Figs. \ref{fig3} display the spin-resolved conductance and spin polarization of the total conductance as functions of the gate voltage $U_g$.

In the weak AM regime, the conductance exhibits spin-dependent oscillations as the gate potential $U_g$ increases from zero [Fig. \ref{fig3}(a)], arising from quantum confinement effects \cite{Cayao2015SNS,Cayao2021Confinement}. 
The conductance peaks appear when the resonance condition [Eq. (\ref{eq_resonant})] is satisfied. 
Between the resonance peaks, one spin channel is strongly suppressed by the tunneling barrier, while the other remains finite, leading to asymmetric spin-resolved conductances.
This asymmetry gives rise to finite spin polarization $P$ [Fig. \ref{fig3}(b)]. 
At low barrier strength (e.g., $V = 0$), the conductance is only partially spin-polarized, whereas at higher barriers (e.g., $V = 10$), a fully spin-polarized conductance is realized with a switchable polarization between $P = +1$ and $P = -1$ by tuning the gate voltage $U_g$.

In the case of strong AM, the conductance as a function of $U_g$ exhibits a highly tunable behavior, persisting even for $U_g<0$, see Fig. \ref{fig3}(c).
This can be understood by the unique Fermi surface geometry of strong AM [Fig. \ref{fig2}(a-iv)], which leads to a regime in which the conductance is exclusively carried by spin-up electrons for $U_g > 0$ and by spin-down electrons for $U_g < 0$. 
As a result, the spin polarization reaches $P = \pm 1$ and can be flipped simply by tuning the gate voltage, as shown in Fig. \ref{fig3}(d). 
Importantly, this spin filtering effect remains robust regardless of the interface transparency set by the tunneling barrier $V$, indicating that it arises from intrinsic bulk properties of the strong AM, rather than interfacial effects in the junction.

Beyond the demonstrated control of spin polarization, the AM-based spin filter offers several key advantages for practical implementation. 
First, because the spin polarization originates from intrinsic Fermi surface geometry rather than interface interference or external magnetic fields, the spin filtering mechanism is expected to be robust against moderate disorder and thermal broadening \cite{Hirohata2020Review,NatureMaterials2022New,DalDin2024Antiferromagnetic,Jungwirth2018Multiple,Baltz2016Antiferromagnetic}. 
Second, the ability to reversibly and continuously tune both the magnitude and sign of $P$ using a single electrostatic gate, without altering material composition or invoking spin–orbit coupling, distinguishes this approach from conventional spintronic platforms \cite{Manchon2015New,Galitski2013Spin}. 
Finally, the momentum-selective spin filtering effect in strong AMs may be leveraged to design functional interfaces with superconductors or topological materials \cite{Zhu2023Topological,Chatterjee2025Interplay,Ghorashi2024Altermagnetic,Li2023Majorana,Li2024Realizing,Tanaka2024Theory} and spin-sensitive Josephson effects \cite{Fukaya2025Josephson,Zhao2025Orientation,Ouassou2023dc,Beenakker2023Phase,Zhang2024Finite,Lu2024Josephson,Sun2025Tunable}. 
These features collectively underline the strong AM as a versatile and scalable building block for spintronic devices.

\subsubsection*{Electrically controlled spin valve based on strong AMs}  \label{sec4b}

The gate-controlled spin polarization flipping in the strong-AM spin filter can be further extended to realize a fully electrically tunable spin valve, a functionality inaccessible in weak AMs considered in this work. 
A conventional spin valve consists of two spin filters connected in series \cite{NatureMaterials2022New,Hirohata2020Review}.
Each spin filtering effect is composed of a magnetic layer, where only electrons with spin aligned to the magnetization contribute to the current. 
High (low) conductance state occurs when the magnetization directions of the two ferromagnetic layers are aligned (opposite), denoted as parallel (antiparallel) configurations, which is switched typically by external magnetic field \cite{Hirohata2020Review,NatureMaterials2022New,DalDin2024Antiferromagnetic,Jungwirth2018Multiple,Baltz2016Antiferromagnetic}.

% Consequently, efficient current flows when the magnetization directions of the two ferromagnetic layers are aligned, while it is suppressed when these directions are opposite, known as the parallel and antiparallel configurations, respectively.
% Parallel-antiparallel switching is typically achieved by applying an external magnetic field, which controls the relative magnetization orientations between layers,
% % The resulting difference in resistance under field reversal gives rise to the giant magnetoresistance effect \cite{Baibich1988GMR,Binasch1989GMR}, 
% which forms the basis for modern electronics applications such as data storage technologies, magnetic random access memory, and magnetic field sensors .

\begin{figure}[!t]
    \centering
    \includegraphics[width=1\linewidth]{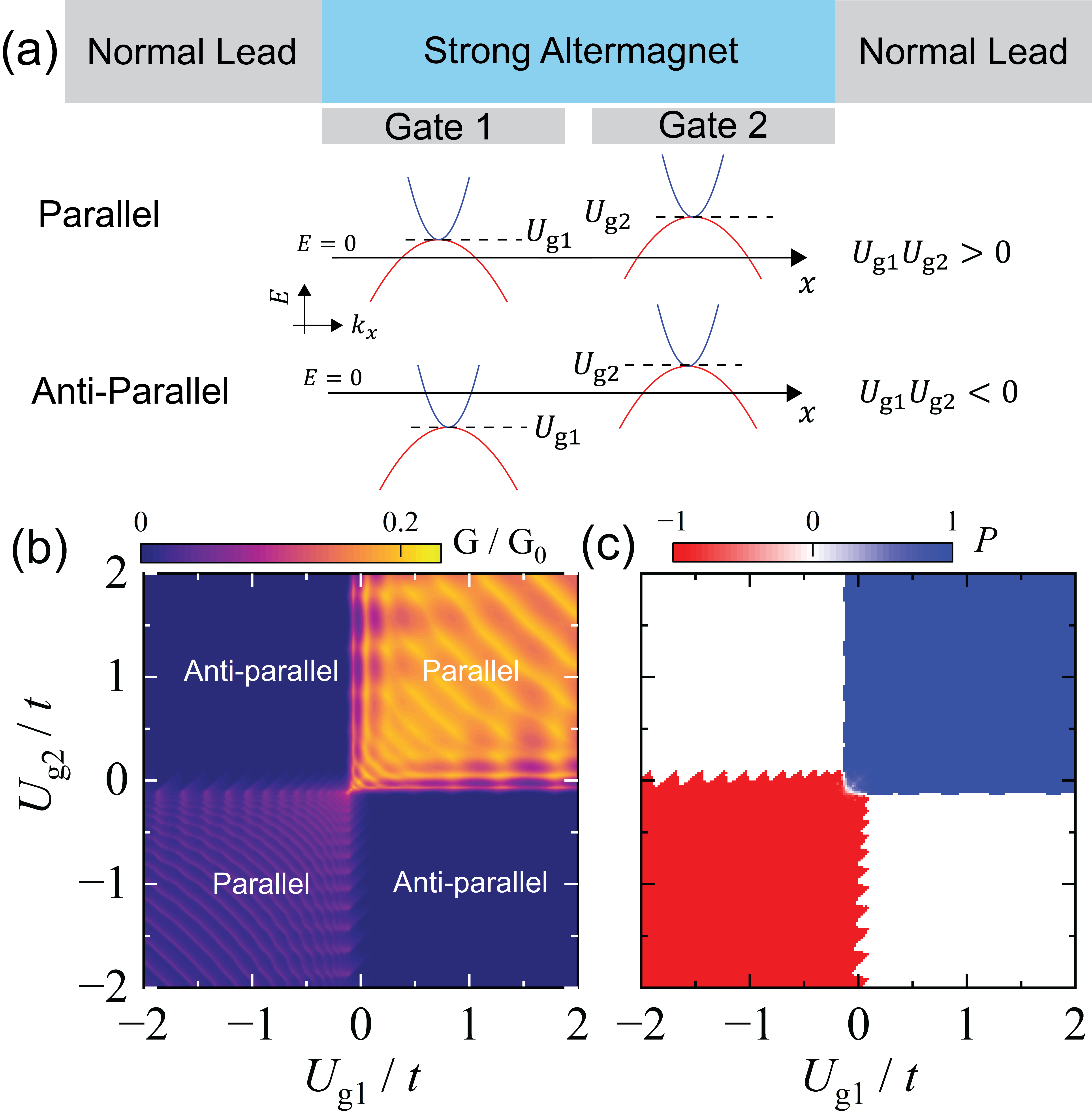}
    \caption{
    (a) Schematic illustration of a double-gate-controlled spin valve based on a strong altermagnetic heterostructure. 
    The configuration enables gate-tunable switching between parallel ($U_{\text{g}1} U_{\text{g}2} > 0$) and antiparallel ($U_{\text{g}1} U_{\text{g}2} < 0$) regimes. 
    (b) Total conductance and (c) spin polarization as functions of the gate voltages $U_{\text{g}1}$ and $U_{\text{g}2}$. 
    The length of the altermagnetic region is $d = 50a$, and the tunneling barrier is set to $V = 0$.
    }
    \label{fig4}
\end{figure}

The gate-tunability in the strong-AM junction enables the realization and manipulation of a spin valve effect with switchable parallel-like and antiparallel-like configurations without net magnetization and external magnetic field. 
The proposed device consists of a strong AM junction with two independently gated regions [Fig. \ref{fig4}(a)], characterized by gate voltages $U_{g1}$ and $U_{g2}$. 
Each gated region serves as a spin filter with spin-up (spin-down) polarization realized by the positive (negative) gate [Figs. \ref{fig3}(c) and (d)]. 
As a result, when both gates are applied in the same direction ($U_{g1} U_{g2} > 0$), spin-polarized conducting channels of identical spin species are activated in both regions [Fig. \ref{fig4}(a)], analogous to a conventional spin valve in the parallel configuration. 
Conversely, when the gate voltages are of opposite sign ($U_{g1} U_{g2} < 0$), different spin-polarized channels are activated in each region [Fig. \ref{fig4}(a)], mimicking the antiparallel configuration of a conventional magnetic spin valve.

Crucially, unlike spin valves composed of ferromagnetic bilayers, where parallel-antiparallel switching requires external magnetic fields to reorient the magnetizations \cite{Hirohata2020Review}, the strong-AM spin valve we proposed is entirely dependent on electrostatic gating. 
The effective spin polarization is governed by the intrinsic momentum-dependent spin-splitting and open Fermi surface of the strong AM [Fig. \ref{fig1}(b)], without requiring net magnetization or magnetic field control. 
This gate-controlled analog of ferromagnetic bilayers establishes strong AMs as a compelling platform for a fully electrical spintronic device.

By solving the scattering problem using the same procedure, the spin-resolved transmission $T_\sigma$ in the spin valve configuration is obtained (see Method). 
Using Eqs. (\ref{eq_Gs})–(\ref{eq_P}), the total conductance $G$ and the resulting spin polarization $P$ are evaluated. 
Their dependence on the gate voltages $U_{g1}$ and $U_{g2}$ is shown in Figs. \ref{fig4}(b) and \ref{fig4}(c), respectively.

Figure \ref{fig4}(b) demonstrates the switching between high and low conductance states by tuning the relative values of the gate voltages $U_{g1}$ and $U_{g2}$ in the spin valve based on a strong AMs [Fig. \ref{fig4}(a)]. 
This effect originates from the opposite spin-dependent band-opening directions in strong AMs [Fig.~\ref{fig4}(a)]. 
As a result, transmission through each gated region is dominated by spin-up (spin-down) electrons for positive (negative) gate voltages, consistent with the spin-filtering behavior shown in Fig. \ref{fig3}(d).
A high-conductance state arises when $U_{g1} U_{g2} > 0$, indicating aligned spin polarizations in the two gated AM regions [Fig. \ref{fig4}(b)]. 
This configuration is analogous to the parallel alignment of magnetizations in conventional ferromagnetic spin valves \cite{NatureMaterials2022New,DalDin2024Antiferromagnetic}. 
More specifically, for $U_{g1}, U_{g2} > 0$ ($< 0$), transport is fully carried by spin-up (spin-down) carriers, yielding a spin polarization of $P = +1$ ($-1$), as shown in Fig. \ref{fig4}(c).
In contrast, when $U_{g1} U_{g2} < 0$, the opposite spin polarization in the two gated regions leads to a strong suppression of conductance due to spin-filter mismatch, resulting in an ill-defined $P$ through Eq. (\ref{eq_P}).
This behavior is reminiscent of the antiparallel configuration in conventional spin valves, where spin-polarized electrons are blocked by the misaligned magnetic orientation of the second layer \cite{NatureMaterials2022New,DalDin2024Antiferromagnetic}.

The gate-tunable switching between high and low conductance states in the strong-AM-based spin valve originates from the opposite spin-dependent band-opening directions [Fig. \ref{fig4}(a)], which is absent in its weak-AM counterpart. 
This functionality arises from electrically controlled spin polarization in each gated strong-AM region, analogous to the spin-filtering behavior in conventional spin valves composed of ferromagnetic bilayers.
However, unlike conventional spin valves, where switching between parallel and antiparallel configurations typically requires external magnetic fields or spin-transfer torques \cite{NatureMaterials2022New,DalDin2024Antiferromagnetic}, our design operates entirely through electrostatic gating. 
This enables faster, more energy-efficient control and facilitates integration with standard semiconductor platforms \cite{Belashchenko2025Giant}. 
Moreover, the pronounced conductance contrast between the two gate configurations, together with the ability to reversibly switch spin polarization, offers a promising route toward nonvolatile spin-based logic and memory elements \cite{Hirohata2020Review,NatureMaterials2022New,DalDin2024Antiferromagnetic,Jungwirth2018Multiple,Baltz2016Antiferromagnetic}. These results underscore the potential of strong AM as a robust platform for scalable, field-free, and all-electric spintronic device architectures without net magnetization.

As a final remark, although the double-gate spin valve is inaccessible in the weak-AM-based junction in the system concerned here.
In recently developed ferroelectric-switchable AMs \cite{Duan2025Antiferroelectric,Gu2025Ferroelectric}, the spin-split Fermi surface can be reoriented by reversing the ferroelectric polarization. 
This effect is analogous to an electric-field-induced rotation of the altermagnetic orientation angle from $\theta_J = 0$ to $\theta_J = \pi/2$ in the present model. 
A gate-controlled spin valve is also conceivable in weak-AM-based junctions. 
Although such systems often involve multiple bands, which lie beyond the scope of the two-band model considered in this work, they also demonstrate the potential spintronics application of the AM-based junctions.

\section*{Discussion}

The results discussed in the previous section are based on the continuum model of altermagnets [Eq.\,(\ref{eq_hsigma1})] and the standard scattering approach (See Methods) for quantum transport at zero temperature. 
This approach offers a transparent way for unveiling the relationship between the emergent spin transport and the geometry of the Fermi surfaces in altermagnet-based devices: this is the key for understanding the realization of highly controllable spin-filter and spin-valve effects in altermagnet systems with open Fermi surfaces. 
The continuum model [Eq.\,(\ref{eq_hsigma1})] provides the simplest framework for realizing open and closed Fermi surfaces by tuning the competition between the kinetic-energy and altermagnetic terms. 
In systems with closed Fermi surfaces [Fig.\,\ref{fig2}(b)], both spin channels contribute to the conductance, resulting in weak spin polarization; in contrast, systems with open Fermi surfaces possess momentum and energy windows where only one spin channel is active [see Fig.\,\ref{fig2}(c)], enabling fully spin-polarized transport. 
Thus, the presence of an open Fermi surface generically ensures the possibility of achieving perfect spin polarization. 

While the continuum model [Eq.\,(\ref{eq_hsigma1})] captures the fundamental mechanism, it does not account for several aspects essential to device architectures based on real materials and their implementation. 
These additional features, which cannot be fully addressed within the continuum description, are discussed within a lattice model analysis, involving the role of more realistic modeling, the effects of saddle points, as well as disorder and finite temperature, see below.

%\subsection*{Numerical simulation using lattice models}

We first stress that the mechanism for fully spin-polarized transport, and related to the open Fermi surface, is not limited to continuum models but also occurs in more realistic lattice models, hence making our results universal, as we explain below. 
To demonstrate the universality of the mechanism behind our findings within the continuum model, we have carried out quantum transport Green's function calculations lattice Green's function technique \cite{Datta1997Electronic,Fu2025Implementation,Fu2022} for three representative lattice models, with the details given in the Supplementary Material, including (i) a spinful square lattice \cite{Datta1997Electronic,Fu2022,Fu2022b}, (ii) a two-sublattice tetragonal model \cite{Roig2024Minimal}, and (iii) a six-band Lieb lattice model \cite{Brekke2023Two}.
This confirms that our conclusions extend beyond the continuum models and connect to real materials based on experimental conditions.

In lattice models, the geometry of the equal-energy surface (open or closed) depends on the chosen energy. 
Thus, open Fermi surfaces can also arise in weak altermagnets. 
Consequently, weak altermagnets can yield stable and perfect spin polarization when the energy or gate voltage selects an open Fermi surface that favors one spin channel. 
This feature, absent in the continuum description [Eq.\,\ref{eq_hsigma1}], parallels the behavior of strong altermagnets, which always host open Fermi surfaces regardless of energy [see e.g. Fig.\,\ref{fig1}(b)]. 
The resulting conductance and spin polarization, therefore, justify our conclusion that open Fermi surfaces in altermagnets generically enable robust spin polarization.
Besides, the lattice models demonstrate that weak AMs can also be beneficial for the spin filter and spin valve effects, which broadens the range of possible materials that can promote our predicted spin transport effects.

Particularly, the spin-filtering effect in the two-sublattice tetragonal model, with parameters guided by density functional theory calculations, connects directly to real materials \cite{Roig2024Minimal} (Sec. I.B in the supplemental material). 
For instance, the two-dimensional weak altermagnetic candidate RuO$_{2}$ \cite{Feng2022} exhibits closed equal-energy surfaces at low energies and open ones at higher energies \cite{Fedchenko2024Observation}. 
Moreover, open equal-energy surfaces have been reported in strong altermagnetic candidates such as MnTe \cite{Reichlova2024,Rial2024}, La$_{2}$O$_{3}$Mn$_{2}$Se$_{2}$, and Ba$_{2}$CaOsO$_{6}$ \cite{Ubiergo2025Atomic}. 
Moreover, the Lieb lattice model further illustrates systems where weak and strong altermagnetic features coexist depending on the gate voltage, providing a natural description for materials such as La$_{2}$CuO$_{4}$ \cite{Brekke2023Two}.

In summary, the lattice-model calculations confirm the fundamental mechanism based on the simple and representative continuum model: stable and nearly perfect spin polarization arises from open Fermi surfaces, whereas closed Fermi surfaces yield unstable and only partially spin-polarized conductance, which is illustrated in Fig.\,\ref{fig2}. 

%\subsection*{Effect of saddle points}

In both the continuum and lattice models, there are several saddle points characterized by the high density of states existing in the altermagnetic systems [See e.g. Fig.\,\ref{fig1}(b) and Sec. I.A in the supplemental material].
However, in the device proposed here, the effect of saddle point(s) on the transport behaviors is negligible.
This can be explained by the tunneling conductance \cite{Bardeen1961Tunnelling,Gray1965Tunneling}, which is proportional to $T(E)$, the transmission probability across the central altermagnetic region, and the density of states, $N_{L(R)}(E)$, in the left (right) normal metal electrodes, i.e.
$G \propto N_{L}(E)N_{R}(E)T(E)$
Thus, the conductance is related to the density of states of the electrodes rather than the altermagnets.
Moreover, saddle points always occur at the extreme value of the dispersion, which implies that at these points $\partial_{k_x}E=0$. 
The transmission probability $T(E)$, related to the velocity [$T(E)\propto v_x \propto \partial_{k_x}E$], thus vanishes around the saddle points.
The influence of saddle points on conductance, and thus on spin polarization, is negligible.

Nevertheless, saddle points act as transition markers between closed and open surfaces in weak altermagnets, or between open surfaces with opposite spins (see Sec. I of the supplemental materials). 
These transition points are reflected in the vanishing conductance of one spin channel and the sharp jumps in spin polarization, see e.g. Fig.\,\ref{fig3} (b) and (d).

Moreover, our work focuses on the gate-tunable spin-dependent transport behaviors. 
This means (i), in equilibrium, Fermi levels are renormalized to be shared in three regions composing the junction (two electrodes and the central altermagnet may have different Fermi levels initially) and determined by the energy of the injecting electrons \cite{Datta1997Electronic}; (ii) the applied gate voltage shifts the energy of the saddle points in the altermagnetic region, which can be closed to or near the Fermi level depending on the gates.

%\subsection*{Effect of disorder}

Introducing the lattice model enables us to examine the robustness of the spin-filtering effect against disorder. 
By incorporating random on-site disorder potentials into the spinful square lattice model (Sec.\,I of the Supplemental Material), we find that the high spin polarization achieved in altermagnets with open Fermi surfaces persists even under strong disorder. 
Remarkably, disorder can also induce perfect spin polarization in weak altermagnets with closed Fermi surfaces, which we attribute to the formation of an effective open Fermi surface.

%\subsection*{Finite temperature effect}

The results above were obtained within the continuum model and quantum tunneling framework at zero temperature. 
The robustness of perfect spin polarization ($P=\pm 1$) in altermagnets with open Fermi surfaces is further embodied in its persistence under finite temperature. 
The spin-resolved conductance at temperature $T$ is given by
\begin{equation}
    G_{\sigma} = G_{0} \int_{-\infty}^{+\infty} \int_{-\pi/2}^{+\pi/2} 
    T_{\sigma}(E,\theta_{k}) 
    \left(-\frac{\partial f}{\partial E}\right) 
    \cos\theta_{k}\, d\theta_{k}\, dE,
    \label{eq_GfiniteT}
\end{equation}
where $f=[1+\exp((E-E_F)/k_{B}T)]^{-1}$ is the Fermi--Dirac distribution. 
Here $G_{0}$ and $T_{\sigma}(E,\theta_{k})$ denote the conductance unit and spin-resolved transmission probability as in Eq.\,(\ref{eq_Gs}). 
At $T=0$, Eq.\,(\ref{eq_GfiniteT}) reduces to Eq.\,(\ref{eq_Gs}) since $-\partial f/\partial E \rightarrow \delta(E-E_F)$. 

\begin{figure*}
    \centering
    \includegraphics[width=1\linewidth]{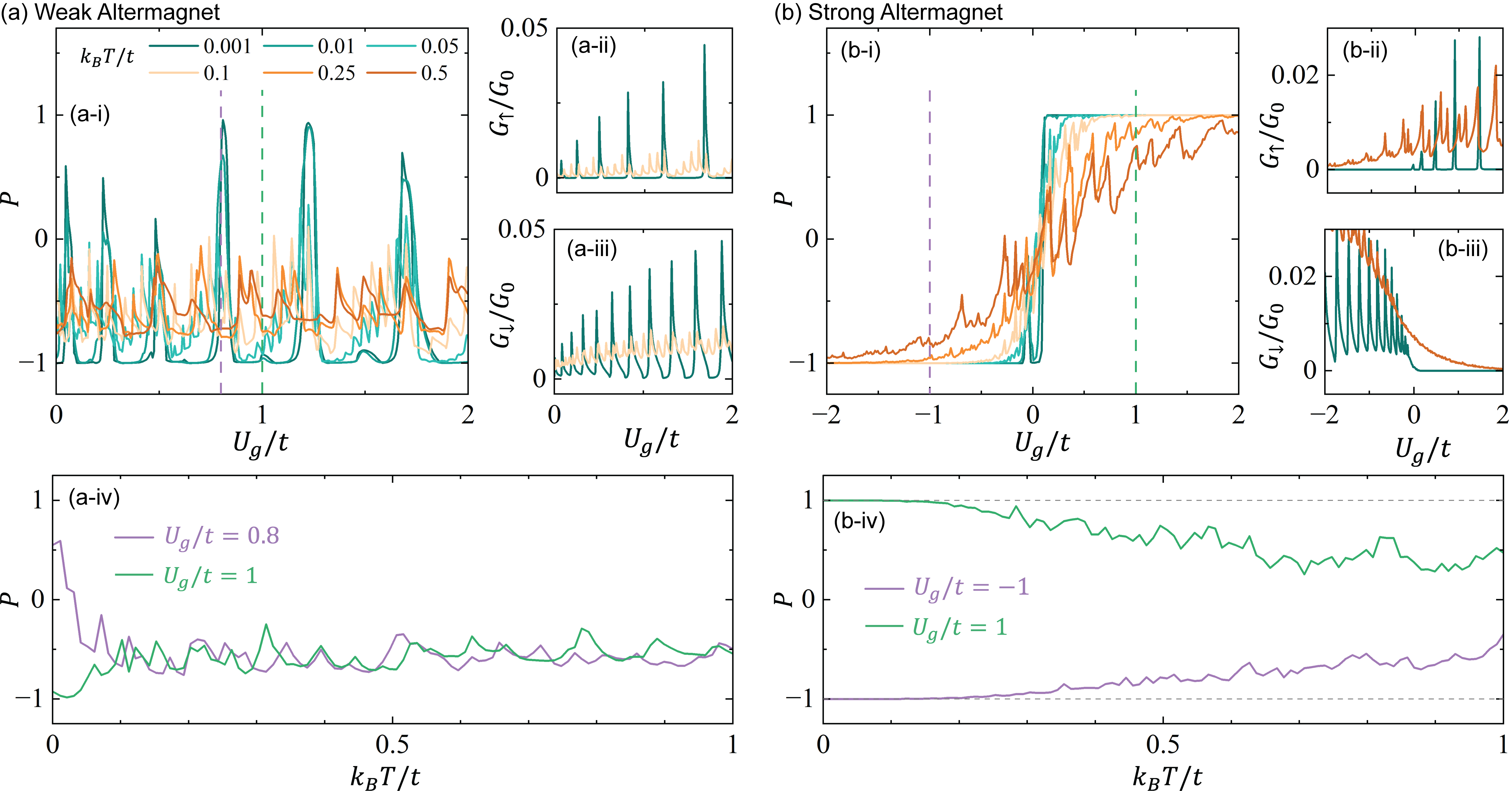}
    \caption{Effect of temperatures $k_BT$ (in the units of $t$) on the conductance [Eq.\,(\ref{eq_GfiniteT})] and spin polarization [Eq.\,(\ref{eq_P})].
    (a-i) Gate-controlled spin polarization for various temperatures $k_BT$ in a weak altermagnetic junction with $J/t=0.5$.
    (a-ii) and (a-iii): Gate-controlled conductance at low and finite temperatures with $k_BT=0.001t$ and $k_BT=0.1t$, respectively.
    (a-iv) Spin polarization as a function of temperatures with specific gate voltage denoted by the vertical dashed lines in (a-i).
    (b) same as (a) but for a strong altermagnetic junction with $J/t=1.5$.
    Parameters: same as Fig.\,\ref{fig3}.
    }
    \label{figureR4}
\end{figure*}

For weak altermagnets with closed Fermi surfaces [Fig.\,\ref{figureR4}(a)], increasing $k_{B}T$ broadens the resonant conductance peaks and partially restores the suppressed background [Fig.\,\ref{figureR4}(a-ii,iii)]. 
This thermal smearing reduces the spin polarization [Fig.\,\ref{figureR4}(a-i)], as expected~\cite{Datta1997Electronic}. 
Nevertheless, finite spin polarization persists up to $k_{B}T\sim 0.01t$, corresponding to $T\sim 11.6$ K for $t\sim 100$ meV. 
In contrast, strong altermagnets with open Fermi surfaces display qualitatively more robust behavior [Fig.\,\ref{figureR4}(b)]. 
Despite thermal broadening, the gate-controlled switching between spin-up and spin-down currents remains sharp, and the polarization stays close to $\pm 1$ up to $k_{B}T \sim 0.25t$, corresponding to $T\sim 290$ K (room temperature). 
This places strong altermagnets with open Fermi surfaces among the most promising candidates for room-temperature spintronics \cite{Moodera1995Large,Jiang2025A}. 
The contrasting thermal stability can be traced to the Fermi-surface geometry. 
In the weak regime, closed Fermi surfaces host two spin channels of nearly equal weight, resulting in partial polarization sensitive to interface transparency [see Fig.\,\ref{fig2}(a,b)]. 
In the strong regime, open Fermi surfaces allow only a single spin channel at a given energy or gate voltage, leading to robust and nearly perfect polarization [see Fig.\,\ref{fig2}(a,c)]. 
These results substantiate our central claim: spin-filtering effects in altermagnets, particularly those with open Fermi surfaces, remain robust under finite-temperature conditions. 
Together with the disorder analysis, they confirm that Fermi-surface geometry is the key factor ensuring stable spin-selective transport in realistic experimental settings.

%\subsection*{Device implementations}

Recent advances in material synthesis have demonstrated that epitaxial growth techniques can stabilize altermagnetism in thin films, paving the way for device integration. 
For example, epitaxial Mn$_5$Si$_3$ on Si(111), grown using MnSi seed layers, exhibits variant-sensitive anomalous Hall anisotropy in nanostructures \cite{Reichlova2024}. 
Similarly, RuO$_2$ thin films under epitaxial strain host altermagnetic phases with ordering temperatures above 500\,K \cite{Dufouleur2023}. 
More recently, single-variant RuO$_2$(101) films grown on $r$-plane Al$_2$O$_3$ substrates have displayed spin-splitting magnetoresistance in CoFeB bilayers, underscoring the decisive role of epitaxial control and variant selection \cite{He2025Evidence}. 
These results highlight the importance of crystalline quality and epitaxial engineering, since both Mn$_5$Si$_3$ and RuO$_2$ exhibit strong sensitivity of their transport signatures to structural quality and domain formation \cite{Reichlova2024}. 
Maintaining a sharp Fermi-surface anisotropic geometry, therefore, requires high crystalline order and careful variant control.

Alongside material growth, interface engineering is crucial for realizing altermagnet-based devices (see e.g. Fig.\ref{fig3}). 
The integration of RuO$_2$ into magnetic tunnel junctions has already produced measurable tunneling magnetoresistance \cite{GonzalezBetancourt2023,Feng2022}, establishing proof of principle for device functionality. 
These observations demonstrate that clean epitaxial interfaces, controlled oxygen stoichiometry, and minimized interfacial disorder are essential for preserving spin-polarized transport across junctions.

A central feature of our proposal is the use of gate-voltage control as a practical and all-electrical means to manipulate spin polarization without external magnetic fields. 
Altermagnets with open Fermi surface naturally support gate-reversible, fully spin-polarized transport that is robust against interface scattering. 
Recent theoretical studies have reinforced this prospect: CrS bilayers, for example, have been predicted to exhibit layer-spin locking, achieving sign-reversible spin polarization of up to $\sim 87\%$ at room temperature under out-of-plane electric fields \cite{Peng2025All}. 
Our results connect directly to these proposals by showing that open Fermi surfaces in altermagnets enable stable, gate-controlled spin filtering, establishing an electrical route to GMR-like functionalities without applied fields and thereby enhancing scalability.

Despite these advances, challenges remain for embedding altermagnets into larger spintronic architectures. 
Spin-memory loss at interfaces can suppress spin polarization during transmission \cite{Rojas-Sanchez2014Spin,Korzhovska2020Spin}, while conductance mismatch between metallic altermagnets and semiconductors remains a well-known obstacle for efficient spin injection \cite{Schmidt2000Fundamental,Awschalom2007Challenges}. 
Furthermore, scalable device fabrication requires reproducible growth of epitaxial thin films with controlled variants and interfaces, as emphasized in recent spintronic roadmaps \cite{Hirohata2020Review,NatureMaterials2022New}. 
Promising strategies to address these limitations are emerging. 
Employing low-resistance altermagnet/normal-metal junctions can mitigate impedance mismatch and minimize spin-loss channels \cite{Rojas-Sanchez2014Spin}. 
Heterostructures combining altermagnets with ferromagnets or topological materials have been proposed theoretically \cite{Smejkal2022a,Smejkal2022,Das2023Transport,Das2024Crossed,Nagae2025Spin}, offering routes to multifunctional spintronic platforms. 
Crucially, the robust spin filtering we demonstrate, rooted in the open Fermi surfaces of strong altermagnets, ensures stable, gate-reversible, and fully spin-polarized transport. 
This intrinsic stability provides a decisive advantage over conventional ferromagnetic systems, where perfect polarization is rarely realized due to strong magnetization. 
The purely electrical tunability of altermagnets makes them naturally compatible with semiconductor-based and CMOS platforms, positioning them as promising building blocks for future large-scale spintronic circuits \cite{Hirohata2020Review,NatureMaterials2022New}.

%\section*{Conclusions} \label{sec5}

We have identified a universal transport mechanism for achieving perfect spin-polarized conductance in altermagnet-based spintronic devices by utilizing the open Fermi surfaces of altermagnets. 
The mechanism is first illustrated in a pedagogical continuum model as a minimal and representative demonstration. 
By analyzing the interplay between the altermagnetic exchange interaction and kinetic energy, we classified altermagnets into weak and strong regimes, corresponding to closed and open spin-resolved Fermi-surface configurations, respectively. 
In the weak regime, the closed Fermi surface hosts two opposite spin channels that contribute nearly equally to transport. 
The resulting conductance is only partially spin-polarized and remains sensitive to interface transparency. 
In the strong regime, the open Fermi surface enables only a single spin channel at a given energy or gate voltage, producing a robust and nearly perfect spin polarization in the conductance. 
Building on the link between perfect spin polarization and open Fermi-surface geometry, we proposed an electrically tunable spin valve. 
In this device, double-gate control enables transitions between parallel and antiparallel spin configurations without requiring magnetic fields or net magnetization, in sharp contrast to conventional ferromagnetic spin valves. 
The functionality derives from the nonrelativistic spin splitting inherent to altermagnets and highlights their potential as a platform for scalable, all-electrical spintronic devices. 

Lattice-model calculations further confirm the fundamental mechanism captured by the continuum model. 
Stable and nearly perfect spin polarization arises whenever transport involves open Fermi surfaces, whereas closed Fermi surfaces yield unstable and only partially polarized conductance. 
Moreover, open Fermi surfaces also emerge in weak altermagnets at certain energies, allowing perfect spin polarization and extending the pool of candidate materials for spin-filter and spin-valve effects. 
Perfect spin polarization enforced by open Fermi-surface topology is additionally shown to be robust against strong disorder and stable up to room temperature, underscoring the experimental feasibility of the proposed devices. 

In conclusion, altermagnets with open Fermi surfaces provide a versatile route to all-electrically controlled perfect spin polarization. 
The results establish open Fermi-surface geometry as the key ingredient for spin-selective transport and pave the way toward the experimental realization of gate-tunable spin filters and spin valves in realistic altermagnetic compounds.

\section*{Methods}

% \subsection{Scattering formalism and spin-polarized conductance} \label{sec3a}

We consider a spin filter junction oriented along $x$ direction, as shown in Fig. \ref{fig2}(a-i). 
This device is composed of a gated AM sandwiched between two normal metal leads. 
The spin filter can be controlled by the gate $U_g$ applied in the AM region and the tunneling barrier at the interfaces between the normal leads and the AM.
The Hamiltonian below models the junction
\begin{eqnarray}
H_{\text{SF}}^{\sigma }\left( x\right) &=& ta^{2}k_{y}^{2} + ta^{2}\left( -i\partial_{x} \right)^{2} - U(x) \label{eq_HSF}\\
&& - \sigma J(x) a^{2}k_{y}^{2} \cos \left( 2\theta_J \right)   \nonumber \\
&& + a^{2}k_{y} \sin \left( 2\theta_J \right) \sigma \left\{ J(x), -i\partial_{x} \right\}  \nonumber \\
&& + a^{2} \cos \left( 2\theta_J \right) \sigma \partial_{x} J(x) \partial_{x}, \nonumber
\end{eqnarray}
where
\begin{eqnarray}
U(x) &=& U_{L} \left[ \Theta(-x) + \Theta(x - d) \right] - U_{g} \Theta(x) \Theta(d - x) \nonumber \\
&& - V ( \delta_{x,0} + \delta_{x,d} ) \text{,}
\end{eqnarray}
describes the electrostatic gating potential applied across the junction, with $\Theta(x)$ the Heaviside step function and $\delta_{x,x_i}$ the Kronecker delta.
Here, $U_L$ sets the chemical potential in the left ($x < 0$) and right ($x > d$) normal leads, $U_g$ denotes the gate voltage applied to the central AM region, and $V$ represents the tunneling barriers at the normal–AM interfaces located at $x = 0$ and $x = d$. 
Here, we assume the chemical potentials in both leads are identical, and the same tunneling barriers are applied at the two interfaces.
The AM occupies the central region $0 < x < d$, with
\begin{equation}
J(x) = J \, \Theta(x) \Theta(d - x),
\end{equation}
and the notation $\left\{ \cdots \right\}$ in the third line of Eq. (\ref{eq_HSF}) indicates the anticommutator, ensuring the Hermiticity of the Hamiltonian \cite{Desai2010Quantum,Fu2025Implementation}.
%As discussed in Sec. \ref{sec2}, weak (strong) altermagnetic spin filters correspond to the case $J < t$ ($J > t$). 
Periodic boundary conditions are assumed along the $y$ direction, so that $k_y$ is conserved throughout the three regions of the junction.

The transport properties of the junction can be obtained by solving the quantum scattering problem \cite{Desai2010Quantum,Fu2025Implementation}. 
For a spin-$\sigma$ incident electron with energy $E$ in the direction $\theta_k$, the eigenfunction is given by $\Psi^{\sigma}(x) = \psi^{\sigma}(x)e^{ik_y y}$, with
\begin{equation}
\psi^{\sigma}(x) = \left\{
\begin{array}{ll}
e^{ik_0 x} + r_\sigma e^{-ik_0 x}, & x \leq 0, \\ [4pt]
c_1 e^{ik_+^\sigma x} + c_2 e^{ik_-^\sigma x}, & 0 \leq x \leq d, \\[4pt]
\tau_\sigma e^{ik_0 x}, & x \geq d,
\end{array}
\right.
\label{eq_eigen}
\end{equation}
where $r_\sigma$ and $\tau_\sigma$ are the reflection and transmission coefficients, respectively, and $c_{1,2}$ are the scattering amplitudes in the central AM region. 
The longitudinal and transverse momentum are defined as 
\begin{equation}
    k_{0} = k_F\cos \theta _{k}\text{,}
\end{equation}
and 
\begin{equation}
k_y = k_F \sin \theta_k\text{,}
\end{equation}
in the normal lead region, with $k_F = \sqrt{ (E + U_L)/(t a^2) }$ being the Fermi wave number of the incident electron. 
The electron injection angle satisfies 
\begin{equation}
\theta_k = \arctan(k_y / k_0)\text{.}
\end{equation}
The longitudinal in the AM region is
\begin{equation}
    k_{\pm }^{\sigma } =\frac{-\sigma Jak_{y}\sin (2\theta _{J})\pm \Omega}{%
at_{\sigma }}\text{,}
\label{eq_kspm}
\end{equation}
with
$
\Omega=\sqrt{(
J^{2}-t^{2}) a^{2}k_{y}^{2}+t_{\sigma }( E-U_{g}) } \text{,}
$
and $t_{\sigma }=t+\sigma J\cos (2\theta _{J})$, which depends on the strength $J$, orientation $\theta_J$, and the gate $U_g$ applied in AM, as well as the incident angle $\theta_k$ and energy $E$ of the injecting electron.
Using the boundary conditions associated with Eq. (\ref{eq_HSF}), we impose
\begin{eqnarray}
\Psi^\sigma(0^-) &=& \Psi^\sigma(0^+), \nonumber \\
\Psi^\sigma(d^-) &=& \Psi^\sigma(d^+), \label{Eq_bdy} \\
- t_{-\sigma} \partial_x \Psi^\sigma(0^+) + t \partial_x \Psi^\sigma(0^-) &=& - V_\sigma \Psi^\sigma(0), \nonumber \\
t_{-\sigma} \partial_x \Psi^\sigma(d^-) - t \partial_x \Psi^\sigma(d^+) &=& - V_\sigma^* \Psi^\sigma(d), \nonumber
\end{eqnarray}
where $t_{-\sigma }=t-\sigma J\cos (2\theta _{J})$ and $V_\sigma = V / a^2 - \sigma i J k_y \sin (2\theta_J)$. 
By substituting the eigenfunction [Eq. (\ref{eq_eigen})] into the boundary condition [Eq. (\ref{Eq_bdy})], the spin-resolved transmission amplitude is then given by
\begin{equation}
\tau_\sigma(E, \theta_k) = 2k_0 \frac{ (t_\sigma / t)(k_+^\sigma - k_-^\sigma) e^{-i d(k_0 - k_+^\sigma - k_-^\sigma)} }{ e^{i d k_-^\sigma} Z^{+}_{+} Z^{-}_{-} - e^{i d k_+^\sigma} Z^{-}_{+} Z^{+}_{-} },
\label{eq_taus}
\end{equation}
with
\begin{equation}
Z^{\alpha}_{\pm} = k_0 + iV / (t a^2) + \alpha \frac{t_{-\sigma } k_{\pm}^\sigma + \sigma J k_y \sin 2\theta_J }{t},
\end{equation}
and $\alpha=\pm1$.
The spin-resolved transmission probability [Eq. \ref{eq_Ts}] can be obtained from Eq. (\ref{eq_taus}).

To explain the transport mechanism within this electrically modulated spin valve shown in Fig. \ref{fig4}(a), we use Eq. (\ref{eq_HSF}) but with an electrostatic potential in AM given by 
\begin{eqnarray}
&&U_{g}\Theta \left( +x\right) \Theta \left( d-x\right)  \\
&\rightarrow &U_{g1}\Theta \left( +x\right) \Theta \left( d/2-x\right)
+U_{g2}\Theta \left( x-d/2\right) \Theta \left( d-x\right) \text{,} 
\nonumber
\end{eqnarray}%
where, for simplicity, we assume that each gated region occupies half of the junction length $d$. 
The spin-resolved transmission $T_\sigma$, total conductance $G$, and spin polarization $P$ can be calculated by solving the quantum scattering problem using the scattering states
\begin{equation}
\psi^\sigma(x) = \left\{
\begin{array}{ll}
e^{ik_0 x} + r_\sigma e^{-ik_0 x}, & x \leq 0, \\ [4pt]
c_1 e^{ik_{1+}^\sigma x} + c_2 e^{ik_{1-}^\sigma x}, & 0 < x \leq d/2, \\[4pt]
c_3 e^{ik_{2+}^\sigma x} + c_4 e^{ik_{2-}^\sigma x}, & d/2 < x \leq d, \\[4pt]
t_\sigma e^{ik_0 x}, & x > d,
\end{array}
\right.
\end{equation}
with
\begin{equation}
k_{1,2\pm}^\sigma = k_\pm^\sigma(U_g \rightarrow U_{g1,2}).
\end{equation}
In addition to the boundary conditions in Eq. (\ref{Eq_bdy}), the wavefunction must also satisfy
\begin{eqnarray}
\Psi^\sigma\left[ (d/2)^+ \right] &=& \Psi^\sigma\left[ (d/2)^- \right], \label{eq_bdy2} \\
\partial_x \Psi^\sigma\left[ (d/2)^+ \right] &=& \partial_x \Psi^\sigma\left[ (d/2)^- \right]. \nonumber
\end{eqnarray}
By solving the scattering problem using the same procedure, the spin-resolved transmission $T_\sigma$ in the spin valve configuration shown in Fig. \ref{fig4}(a) is obtained. 
Using Eqs. (\ref{eq_Gs})–(\ref{eq_P}), the total conductance $G$ and the resulting spin polarization $P$ are evaluated. 
Their dependence on the gate voltages $U_{g1}$ and $U_{g2}$ is shown in Figs. \ref{fig4}(b) and \ref{fig4}(c), respectively.

\section*{Data Availability}
The data supporting the findings of this study are available from the corresponding authors upon reasonable request.

\section*{Acknowledgements}
P.-H. Fu appreciates the support from W. Xu and the discussion from K. W. Lee and Z. Yuan. 
Q. Lv acknowledges financial support from Shenzhen University of Information Technology (Grant No. SZIIT2025KJ066).
Y. Xu acknowledges financial support from the Scientific Research Starting Foundation of Ningbo University of Technology (Grant No. 2022KQ51) and the China Postdoctoral Science Foundation (Grant No. 2023M743783).
J. C. acknowledges financial support from the Carl Trygger’s Foundation (Grant No. 22: 2093), the Sweden-Japan Foundation (Grant No. BA24-0003), the G\"{o}ran Gustafsson Foundation (Grant No. 2216), and the Swedish Research Council (Vetenskapsr\aa det Grant No.~2021-04121).
J.-F. L. acknowledges financial support from the National Natural Science Foundation of China (Grant No. 12174077) and the Joint Fund with Guangzhou Municipality under Grant No. 202201020238.
X.-L.Yu acknowledges financial support from the Guangdong Basic and Applied Basic Research Foundation (Grant No. 2023A1515011852) and the Shenzhen Natural Science Foundation (Grant No. JCYJ20250604174400001).

\section*{Author Contributions}
P.-H. F. and Q. L. contributed equally to this work. P.-H. F. conceived the idea, developed the theoretical model, performed calculations, and wrote the manuscript. Q. L. also performed calculations and contributed to manuscript writing. X. Y. and J. C. provided valuable insights and contributed to the analysis and interpretation of the results. X.-L. Y. and J.-F. L. supervised the project. All authors discussed the results and contributed to the final version of the manuscript.

\section*{Competing Interests}
The authors declare no competing interests.

\end{document}

\title{Supplemental Material for "All-electrically controlled spintronics in altermagnetic heterostructures"}

\author{Pei-Hao Fu}
\email{phy.phfu@gmail.com}
\affiliation{School of Science, Sun Yat-sen University, Shenzhen 518107, China}
\affiliation{School of Physics and Materials Science, Guangzhou University, Guangzhou 510006, China}

\author{Qianqian Lv}
%\email{2024000090@sziit.edu.cn}
\affiliation{School of Humanities and Basic Sciences, Shenzhen Institute of Information Technology, Shenzhen 518172, China}

\author{Yong Xu}
%\email{xuyong_nbut@163.com}
\affiliation{Institute of Materials, Ningbo University of Technology, Ningbo 315016, China}

\author{Jorge Cayao}
%\email{jorge.cayao@physics.uu.se}
\affiliation{Department of Physics and Astronomy, Uppsala University, Box 516, S-751 20 Uppsala, Sweden}

\author{Jun-Feng Liu}
\email{phjfliu@gzhu.edu.cn}
\affiliation{School of Physics and Materials Science, Guangzhou University, Guangzhou 510006, China}

\author{Xiang-Long Yu}
\email{yuxlong6@mail.sysu.edu.cn}
\affiliation{School of Science, Sun Yat-sen University, Shenzhen 518107, China}

\date{\today}
\maketitle
\tableofcontents
\newpage
%%%%%%%%%%%%%%%%%%%%%%%%%%%%%%%%%%%%%%%%%%%%%%%%%%%%%%%%%%%%%
\section{Numerical simulation using lattice models}
The continuum models analyzed in the main text serve as a minimal and representative demonstration of this mechanism. 
In the weak altermagnetic regime, the closed Fermi surface hosts two opposite spin channels, yielding nearly equal conductance for both spins and thus a partially spin-polarized conductance sensitive to the interface transparency [see Fig.\,4(a,b) in the main text]. 
By contrast, in the strong altermagnetic regime, the open Fermi surface admits only a single spin channel at a given energy or gate voltage, leading to a robust and nearly perfect spin polarization in the conductance [see Fig.\,4(c,d) in the main text].
Thus, the key to the reported effects is the geometry of the Fermi surfaces.

\begin{figure}
    \centering
    \includegraphics[width=1\linewidth]{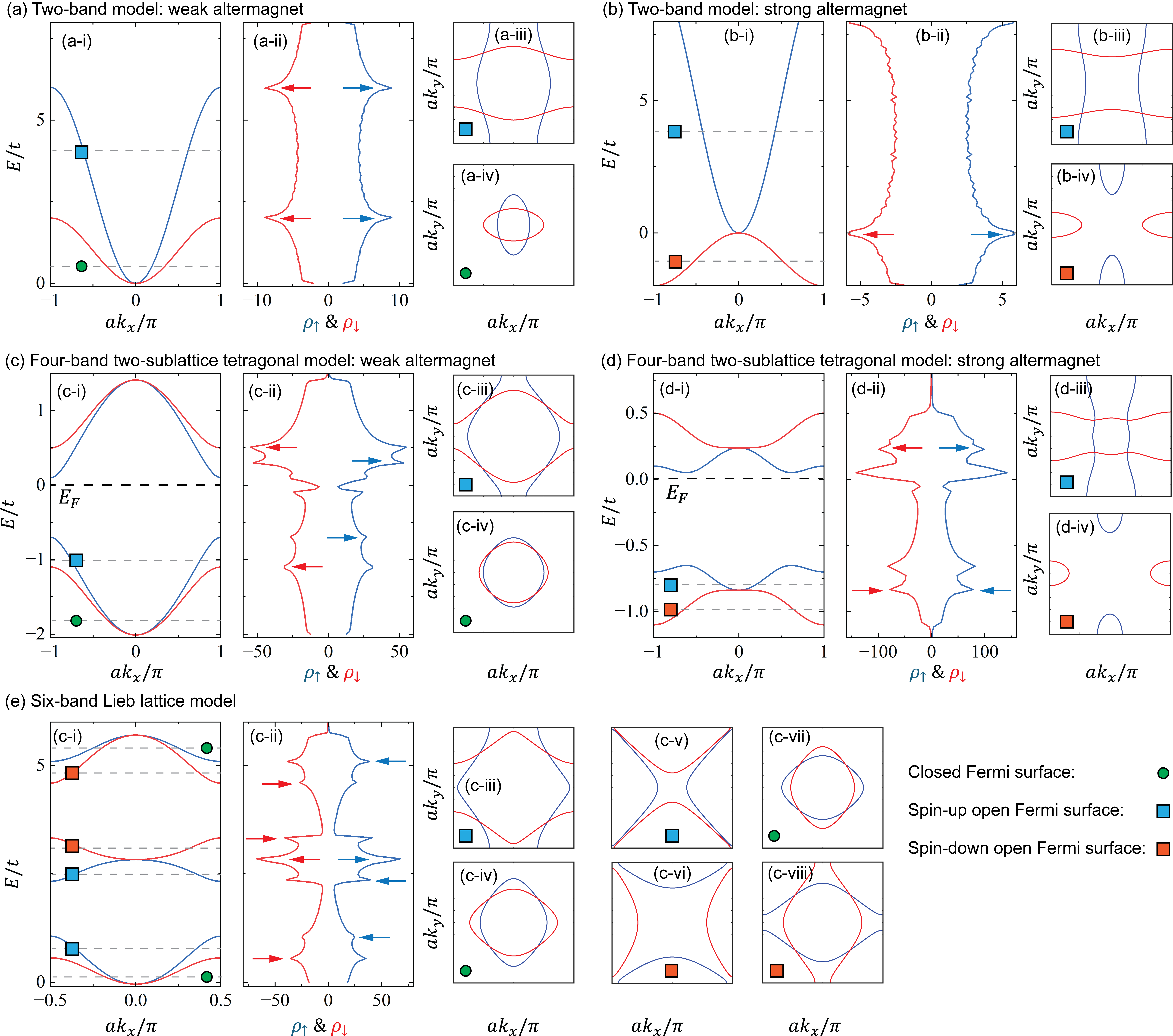}
    \caption{Dispersion, spin-resolved density of states, and Fermi surfaces geometry of various lattice models. 
    (a) Two-band spinful square lattice model [Eq.\,(\ref{eq_hssl})] with weak altermagnetism $J/t=0.5<1$. 
    (a-i) Dispersion along $k_x$ axis with $k_y=0$.
    (a-ii) Spin-resolved density of states [Eq.\,(\ref{eq_rhos})]. The arrows indicate the van Hove singularity (saddle points)
    (a-iii) and (a-iv):  Open and closed equal-energy surfaces in $k_x$-$k_y$ plane corresponding to the horizontal gray dashed line in (a-i), respectively.
    (b) same as (a) but with strong altermagnetism $J/t=1.5>1$. 
    Parameters: $t=1$, $a=1$ and $\theta_J=0$.
    (c) and (d): for the four-band two-sublattice tetragonal model [Eq.\,(\ref{eq_h2t})] with (c) $t_3=1.7$ and (d) $t_3=0.5$, respectively. 
    Parameters: $t_1=-0.1$, $t_2=0.1$, $t_4=0$, $t_4^{\prime}$, $\mu=0$ and $J_z=0.2$. 
    These parameters are guided by the DFT calculations \cite{Roig2024Minimal}.
    The black dashed lines indicate the Fermi level at $E_F=0$.
    In (c-iii), (c-iv), (d-iii), and (d-iv), only the equal-energy surfaces of the lower two bands are exhibited, which are similar to the ones of the upper two bands.
    (e): for the six-band Lieb lattice model [Eq.\,(\ref{eq_h6L})]
    which exhibit strong and weak altermagnetism depending on the energy. 
    Parameters: $t_2/t=0$, $\mu/t=0$, $\epsilon_{nm}/t=0$ and $J/t=0.4$ \cite{Brekke2023Two}
    }
    \label{figureR1}
\end{figure}

We have verified that the mechanism and spin transport explained in continuum models remain valid in lattice models. 
To demonstrate this, we have considered three realistic representative lattice models for altermagnets and carried out extensive quantum transport calculations based on Green's functions. In particular, we have considered the following lattice models: (i) a spinful square lattice [Eq.\,(\ref{eq_hssltb})], (ii) a two-sublattice tetragonal model [Eq.\,(\ref{eq_h2t})], and (iii) a six-band Lieb lattice model [Eq.\,(\ref{eq_h6Ltb})]. 
The calculated conductance and spin polarization, which are of central relevance for spintronic devices, are shown in Fig.\,\ref{figureR2}, while the underlying mechanisms are clarified through the band dispersions and equal-energy surface geometries presented in Fig.\,\ref{figureR1}. 
From these calculations, several Remarks can be drawn:

\textbf{a}. 
In lattice models, the geometry of the equal-energy surface (open or closed) depends on the chosen energy. 
Thus, open Fermi surfaces can also arise in weak altermagnets [see Fig.\,\ref{figureR1}(a-i, a-iii) and (c-i, c-iii)]. 
Consequently, weak altermagnets can yield stable and perfect spin polarization when the energy or gate voltage selects an open Fermi surface that favors one spin channel, see Fig.\,\ref{figureR2}(a) and (c). 
This feature, absent in the continuum description, parallels the behavior of strong altermagnets, which always host open Fermi surfaces regardless of energy [see Fig.\,\ref{figureR2}(b,d)]. 
The results in Fig.\,\ref{figureR2} therefore justify our conclusion that open Fermi surfaces in altermagnets generically enable robust spin polarization, broadening the range of material candidates for altermagnet-based spintronic devices.

\textbf{b}. The influence of saddle points on conductance, and thus on spin polarization, is negligible. 
Nevertheless, saddle points act as transition markers between closed and open surfaces in weak altermagnets [see Fig.\,\ref{figureR1}(a,c) and the upper/lower bands in Fig.\,\ref{figureR1}(e)], or between open surfaces with opposite spins [see Fig.\,\ref{figureR1}(b,d) and the middle bands in Fig.\,\ref{figureR1}(e)]. 
These transition points are reflected in the vanishing conductance of one spin channel and the sharp jumps in spin polarization observed in Fig.\,\ref{figureR2}.

\textbf{c}. The spin-filtering effect in the two-sublattice tetragonal model [Eq.\,(\ref{eq_h2t})] connects directly to real materials. 
For instance, the two-dimensional weak altermagnetic candidate RuO$_{2}$ \cite{Feng2022} exhibits closed equal-energy surfaces at low energies and open ones at higher energies \cite{Fedchenko2024Observation}. 
Moreover, open equal-energy surfaces have been reported in strong altermagnetic candidates such as MnTe \cite{Reichlova2024,Rial2024}, La$_{2}$O$_{3}$Mn$_{2}$Se$_{2}$, and Ba$_{2}$CaOsO$_{6}$ \cite{Ubiergo2025Atomic}. 
The Lieb lattice model [Eq.\,(\ref{eq_h6Ltb})] further illustrates systems where weak and strong altermagnetic features coexist depending on the gate voltage, providing a natural description for materials such as La$_{2}$CuO$_{4}$ \cite{Brekke2023Two}. 

In summary, the lattice-model calculations confirm the fundamental mechanism proposed in the main text on the basis of the simple and representative continuum model: stable and nearly perfect spin polarization arises from open Fermi surfaces, whereas closed Fermi surfaces yield unstable and only partially spin-polarized conductance. 

In the following, to substantiate this conclusion, we analyze three representative lattice models: (i) a spinful square lattice, (ii) a two-sublattice tetragonal model, and (iii) a six-band Lieb lattice model. 
The resulting conductance and spin polarization, relevant for spintronic applications, are presented in Fig.\,\ref{figureR2}, while the underlying mechanisms are clarified through the dispersions and equal-energy surface geometries shown in Fig.\,\ref{figureR1}.

\subsection{spinful square lattice}
A natural starting point is to construct a lattice Hamiltonian corresponding to the continuum model analyzed in the main text. 
Following the standard substitution procedure \cite{Datta1997Electronic,Fu2022,Fu2022b}, we replace (for $i = x,y$)
\begin{eqnarray}
    k_{i} &\rightarrow & \frac{1}{a}\sin (ak_{i}), \\
    k_{i}^{2} &\rightarrow & \frac{2}{a^{2}}\left[ 1-\cos (ak_{i})\right],
\end{eqnarray}%
in Eq.\,(1-4) of the main text, and obtain
\begin{equation}
H_{\text{ssl}}(\bm{k}) = \xi_{\bm{k}}^{\prime}\sigma_{0}
+ J_{\bm{k}}^{\prime}\sigma_{z},  
\label{eq_hssl}
\end{equation}%
with
\begin{eqnarray}
\xi_{\bm{k}}^{\prime} &=& 4t - \mu - 2t\cos (ak_{x}) - 2t\cos (ak_{y}), \\
J_{\bm{k}}^{\prime} &=& J\left[ 2\cos (2\theta_{J})(\cos (ak_{y}) - \cos (ak_{x})) 
+ 2\sin (2\theta_{J})\sin (ak_{x})\sin (ak_{y}) \right].
\end{eqnarray}
Around the $\Gamma$ point, the continuum limit is recovered. 
The eigenvalues of Hamiltonian [Eq.\,(\ref{eq_hssl})] yield the band dispersions shown in Fig.\,\ref{figureR1}(a-i) and \ref{figureR1}(b-i) for weak and strong altermagnets with $J/t<1$ and $J/t>1$, respectively. 
For energies near $\Gamma$, the weak altermagnet exhibits a closed equal-energy surface [Fig.\,\ref{figureR1}(a-iv)], whereas the strong altermagnet shows open equal-energy surfaces [Fig.\,\ref{figureR1}(b-iii,b-iv)]. 
In addition, a saddle point appears at $\Gamma$ in the strong regime. 
As illustrated in Fig.\,\ref{figureR1}(b-i) along $k_{y}=0$, the equal-energy surface corresponds to spin-up (spin-down) states for energies below (above) the saddle point. 
These features, (i) distinct equal-energy surfaces in weak versus strong altermagnets and (ii) spin-polarization reversal across the saddle point, are captured by the continuum model, as demonstrated in Fig.\,2 of the main text.
\begin{figure}
    \centering
    \includegraphics[width=0.9\linewidth]{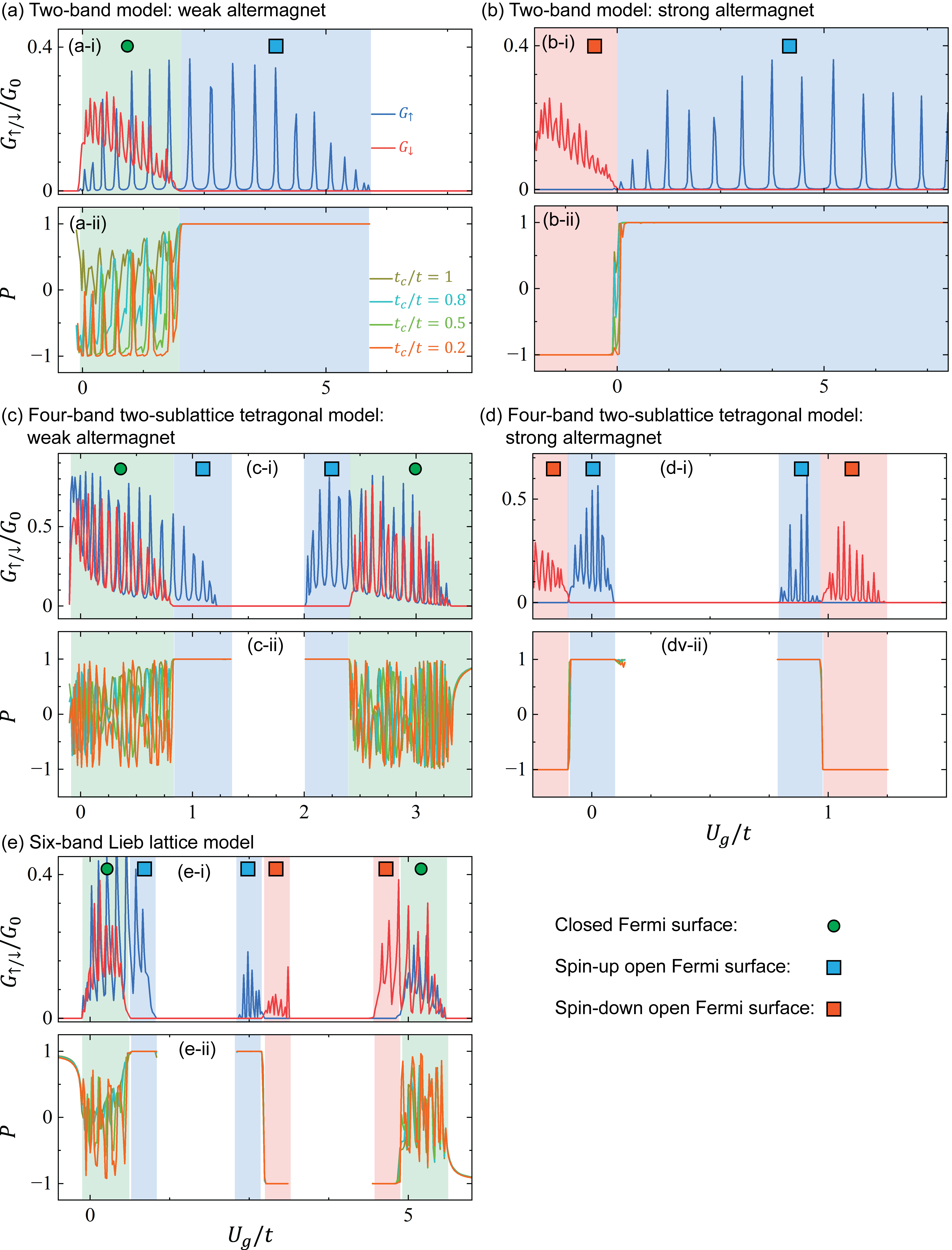}
    \caption{Gate-controlled conductance [Eq.\,\ref{eq_G}] and spin polarization [Eq.\,\ref{eq_P}] obtained from lattice models corresponding to Fig.\,\ref{figureR1}.
    The blue and red shaded regions denote the perfectly spin-polarized conductance contributed by the spin-up and spin-down open Fermi surfaces.
    The green shaded regions denote the partially spin-polarized conductance due to the closed Fermi surfaces.
    Parameters: same as Fig.\,\ref{figureR1}.
    }
    \label{figureR2}
\end{figure}

At higher energies, the continuum description becomes invalid. 
In this regime, the weak altermagnet can also host open equal-energy surfaces [see Fig.\,\ref{figureR1}(a-iii)], whereas in the strong altermagnet the equal-energy surfaces remain open for any energy [see Fig.\,\ref{figureR1}(b-iii,b-iv)]. 
Importantly, the open equal-energy surfaces in the weak altermagnet \textbf{cannot} be captured within the continuum model, and their appearance is always accompanied by saddle points. 
% This observation is fully consistent with the referee’s remark that saddle points are ubiquitous in band dispersions, whether magnetic or nonmagnetic, altermagnetic or ferromagnetic ({\color{blue} Comment 1.2}).

The transitions between closed and open equal-energy surfaces in weak altermagnets, and between open surfaces with opposite spin polarization in strong altermagnets, are associated with the appearance of saddle points at the Brillouin-zone boundaries. 
To reveal these features, we evaluate the spin-resolved density of states (DOS) using the Green’s function corresponding to Eq.\,(\ref{eq_hssl}):
\begin{equation}
\rho_{\uparrow(\downarrow)}(E) = -\frac{1}{\pi}\int d^{2}\bm{k}\;
\mathrm{Tr}\,\mathrm{Im}\left[ \sigma_{\uparrow(\downarrow)} g(E)\right],
\label{eq_rhos}
\end{equation}
with
\begin{equation}
g(\bm{k},E) = \left[E+i\eta - H_{\text{ssl}}(\bm{k})\right]^{-1},
\end{equation}
the Green's function associated with Eq.\,(\ref{eq_hssl}),
where $\eta=10^{-3}$ is a small broadening and 
$\sigma_{\uparrow(\downarrow)} = [\sigma_{0} \pm \sigma_{z}]/2$ are the spin projectors. 

Fig.\,\ref{figureR1}(a-ii) shows $\rho_{\uparrow(\downarrow)}(E)$ for the weak altermagnet, which exhibits peaks at 
$E_{\text{edge}}^{\sigma} = 4t - \mu + 4\sigma J$. 
These peaks correspond to the saddle points at the Brillouin-zone edges, $(k_{x},k_{y})=(0,\pm\pi)$ or $(\pm\pi,0)$ [see Fig.\,\ref{figureR1}(a-i)]. 
For $E < E_{\text{edge}}^{-}$, the equal-energy surface is closed [Fig.\,\ref{figureR1}(a-iv)], while for $E_{\text{edge}}^{-}<E<E_{\text{edge}}^{+}$ it becomes open, and for $E > E_{\text{edge}}^{+}$ the surface closes again. 
Thus, the closed-to-open transition in the weak altermagnet is marked by the saddle-point singularities in the DOS. 

In the strong altermagnet, the equal-energy surface is open throughout, but saddle points appear at $\Gamma$ ($E_{\Gamma}=0$) and at the Brillouin-zone corners [$E_{\text{corner}}=8t$ at $(k_{x},k_{y})=(\pm\pi,\pm\pi)$]. 
These features correspond to the reversal of spin polarization in states along $k_{y}=0$, as illustrated in Fig.\,\ref{figureR1}(b-iii,b-iv).

Although saddle points exist in both weak and strong altermagnets, their effect on transport is negligible. 
The essential mechanism governing the distinct spin-transport behaviors is therefore determined by the geometry of the Fermi surface (open versus closed), rather than the presence of saddle points. 

To analyze the conductance of the junction proposed in this work, we Fourier transform Eq.\,(\ref{eq_hssl}) and obtain the tight-binding Hamiltonian of the altermagnet,
\begin{equation}
H_{\text{ssl}}^{\text{tb}} = \sum_{i,j} h_{0} C_{i,j}^{\dagger} C_{i,j} 
+ h_{x} C_{i+1,j}^{\dagger} C_{i,j} 
+ h_{y} C_{i,j+1}^{\dagger} C_{i,j} 
+ h_{+} C_{i+1,j+1}^{\dagger} C_{i,j} 
+ h_{-} C_{i+1,j-1}^{\dagger} C_{i,j} 
+ h.c., 
\label{eq_hssltb}
\end{equation}
with
\begin{eqnarray}
h_{0} &=& (4t-U_{g})\,\sigma_{0}, \\
h_{x(y)} &=& -t\sigma_{0} \mp J\cos(2\theta_{J})\,\sigma_{z}, \\
h_{\pm} &=& \mp \frac{J}{2}\sin(2\theta_{J})\,\sigma_{z}.
\end{eqnarray}

The tight-binding Hamiltonian of the normal-metal leads, $H_{\text{N}}$, follows by setting $J=0$ and $U_{g}=U_{L}$ ($U_{L}$ is the chemical potential in the leads) in Eq.\,(\ref{eq_hssltb}). 
The total junction Hamiltonian is then
\begin{equation}
H_{\text{junc}} = H_{\text{ssl}}^{\text{tb}} + H_{\text{N}} + H_{\text{C}}, 
\label{eq_junc}
\end{equation}
where the coupling term  reads
\begin{equation}
H_{\text{C}} = -\sum_{j} t_{c}\,\sigma_{0}\, C_{i+1,j}^{\dagger}C_{i,j} + h.c.,
\end{equation}
and $t_{c}$ describes the transparency of the interface at $x=0$ and $x=d$ where $d$ is the length of the junction.
For $t_{c}=1$ ($t_{c}\ll1$), the junction is in the ballistic (tunneling) limit with low (high) tunneling barrier. 
At zero temperature, the conductance is expressed in terms of the total transmission probability $T(k_{y},E)$ as \cite{Datta1997Electronic,Fu2022}
\begin{equation}
G = G_{\uparrow}+G_{\downarrow} 
= \frac{e^{2}}{h} \sum_{k_{y},\,\sigma=\uparrow,\downarrow} 
T_{\sigma}(k_{y},E), 
\label{eq_G}
\end{equation}
where translational invariance along $y$ ensures that $k_{y}$ is conserved. 
The spin-resolved transmission probability is
\begin{equation}
T_{\sigma}(k_{y},E) = 
\mathrm{Tr}\left[ \Gamma_{L}^{\sigma} G_{LR}^{\sigma,r} 
\Gamma_{R}^{\sigma} G_{LR}^{\sigma,a}\right],
\end{equation}
with $\Gamma_{L/R}^{\sigma}=i(\Sigma_{L/R}^{\sigma,r}-\Sigma_{L/R}^{\sigma,a})$ the linewidth function, $\Sigma_{L/R}^{\sigma,r}$ the self-energies from the NM leads, and $G_{LR}^{\sigma,r}=(E-H_{\text{ssl}}^{\text{tb}}-\Sigma_{L}^{r}-\Sigma_{R}^{r})_{\sigma}^{-1}$ the retarded Green’s function, obtained via the lattice Green’s function technique \cite{Fu2022,Fu2025Implementation}. 
Here $\sigma=\uparrow,\downarrow$ denotes the spin block of the matrices.  
The spin polarization of the conductance is defined as
\begin{equation}
P = \frac{G_{\uparrow}-G_{\downarrow}}{G_{\uparrow}+G_{\downarrow}}, 
\label{eq_P}
\end{equation}
which coincides with Eq.\,(11) of the main text.

%Calculations based on the lattice model confirm the transport mechanism proposed in this work: systems with open Fermi surfaces exhibit stable and perfect spin polarization ($P=\pm 1$). 

The conductance and spin polarization as functions of the gate voltage $U_{g}$ are shown in Fig.\,\ref{figureR2}(a,b) for weak and strong altermagnets. 
For small $U_{g}$, the weak altermagnet hosts a closed Fermi surface, resulting in oscillating spin polarization in the tunneling limit [see Fig.\,\ref{figureR2}(a-ii)]. 
This behavior arises from different resonance conditions between spin-up and spin-down channels [see Fig.\,\ref{figureR2}(a-i) and Eq.\,(13) in the main text]. 
At higher $U_{g}$, the spin-up open Fermi surface of the weak altermagnet [Fig.\,\ref{figureR1}(a-iii)] dominates transport, while the spin-down channel is completely suppressed. 
As a result, the conductance becomes perfectly spin polarized ($P=+1$), robust against variations in interface transparency (captured by different $t_{c}$). 
In the strong altermagnet, perfect and stable spin polarization is also observed [see Fig.\,\ref{figureR2}(b)]. 
For $U_{g}<0$ ($U_{g}>0$), only the spin-down (spin-up) open Fermi surface contributes to transport, so the conductance is carried solely by a single spin channel, yielding $P=-1$ ($P=+1$). 
Although the contribution of saddle points to conductance is negligible, they signal geometric transitions of the Fermi surface. 
This is manifested in the critical gate voltages of Fig.\,\ref{figureR2}(a-ii,b-ii), which mark the change from unstable to stable spin polarization in weak altermagnets, and the reversal of spin polarization in strong altermagnets. 

From Fig.\,\ref{figureR2}(a,b) we conclude that whenever an open Fermi surface is present, perfect and stable spin polarization can be realized in altermagnetic junctions, irrespective of whether the system is in the weak or strong regime. 
The weak and strong altermagnets studied in the manuscript provide the simplest cases of closed and open Fermi surfaces, respectively, and thus capture the essential physics. 
In the following, we demonstrate two models closely connected to real materials, further justifying the mechanism by which open Fermi surfaces in altermagnets yield stable and perfect spin polarization.

\subsubsection{Effect of disorder}
\begin{figure}
    \centering
    \includegraphics[width=1\linewidth]{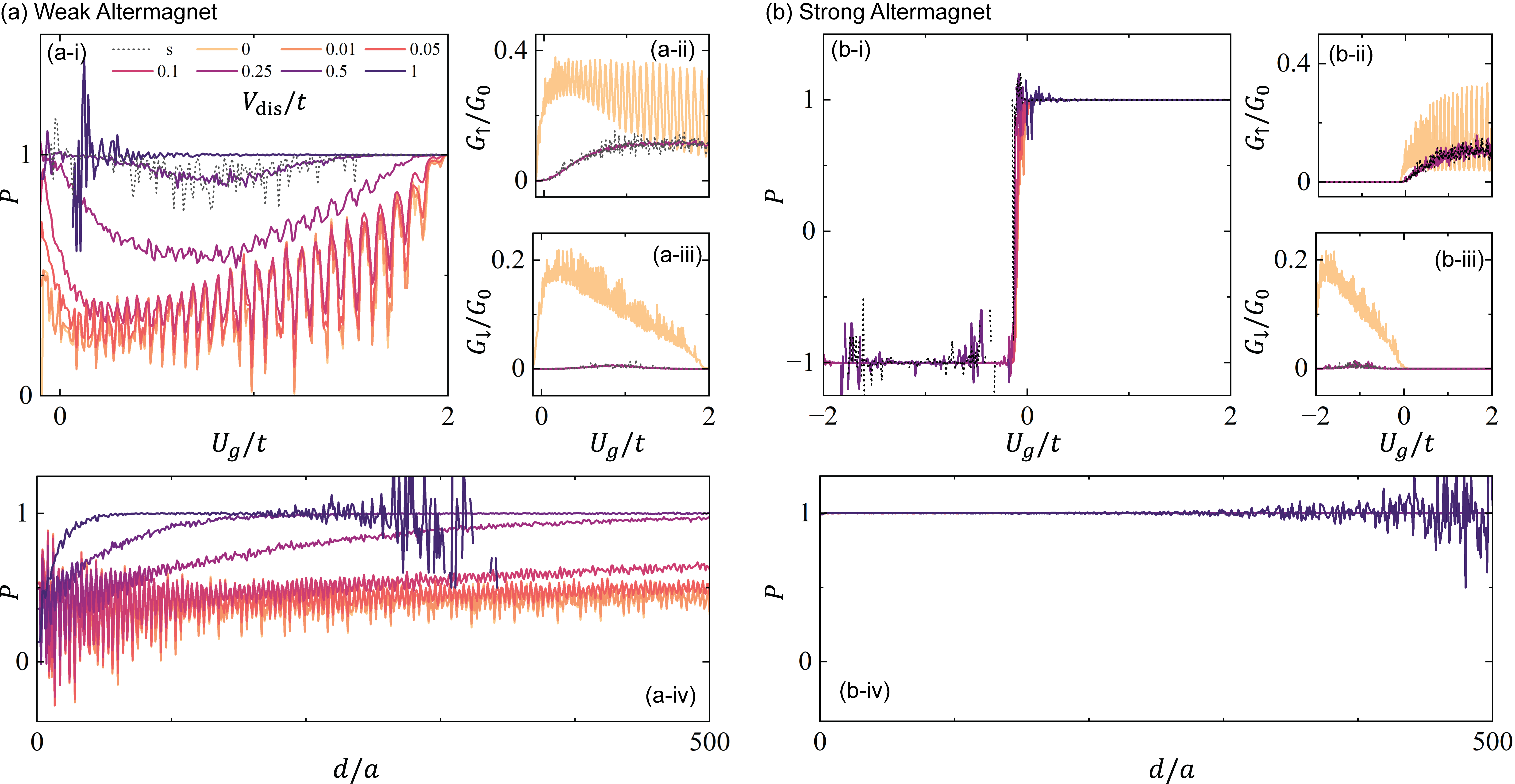}
    \caption{Effect of disorder $V_{\text{dis}}$ (in units of $t$) on the conductance [Eq.~(\ref{eq_G})] and spin polarization [Eq.~(\ref{eq_P})]. 
    All results are averaged over 50 independent disorder realizations, unless otherwise noted.  
    (a) Weak altermagnet ($J/t=0.5$):  
    (a-i) Gate-controlled spin polarization $P$ for various disorder strengths.  
    The black dotted line labeled "s" shows a single realization at $V_{\text{dis}}/t=0.5$ before ensemble averaging.  
    (a-ii) and (a-iii) Gate-controlled conductance for spin-up and spin-down channels at $V_{\text{dis}}=0$ and $V_{\text{dis}}/t=0.5$, respectively. The junction length is $d=100$.
    (a-iv) Spin polarization as a function of the junction length $d$ (in units of the lattice constant $a$) at a fixed gate voltage $U_g/t=1$.  
    (b) Strong altermagnet ($J/t=1.5$): same quantities as in (a). 
    Parameters are the same as in Fig.\,\ref{figureR2}(a) and (b).}
    \label{figureR3}
\end{figure}

Disorder and imperfections are modeled by adding random on-site potentials to the lattice Hamiltonian [e.g. Eq.\,\ref{eq_hssltb}],  
\begin{equation}
H_{\text{dis}} = \sum_{i,j} v_{\text{dis}} \, 
C^{\dagger}_{i,j} C_{i,j}, 
\label{eq_Hdis}
\end{equation}
where $v_{\text{dis}}$ is the strength of the on-site non-magnetic disorder, uniformly distributed within the interval $[-V_{\text{dis}}, V_{\text{dis}}]$. 
Following the procedure from Eqs.\,(\ref{eq_G})--(\ref{eq_P}), we calculate the disorder-averaged conductance and spin polarization, with the results shown in Fig.\,\ref{figureR3}. 

Interestingly, disorder can greatly enhance the spin polarization of altermagnets with closed Fermi surfaces. 
As shown in Fig.\,\ref{figureR3}(a) for the weak altermagnet ($J/t=0.5$), in the low-gate-voltage regime where the closed Fermi surface is engaged [see Fig.\,\ref{figureR1}(a)], the disorder-free junction ($V_{\text{dis}}=0$) exhibits only partial spin polarization because both spin channels contribute comparably to the transport [see Fig.\,\ref{figureR3}(a-ii) and (a-iii)]. 
In this regime, fully spin-down-polarized transport ($P=-1$) appears only in the tunneling limit, while spin-up polarization emerges only at higher gate voltages [see Fig.\,\ref{figureR2}(a)]. 
Surprisingly, increasing $V_{\text{dis}}$ suppresses both spin channels but in an asymmetric manner: the spin-down conductance is strongly reduced while the spin-up conductance remains finite, leading to nearly perfect spin-up polarization $P=+1$ [see Fig.~\ref{figureR3}(a-ii) and (a-iii)]. 
This behavior can be understood perturbatively: disorder enters the Hamiltonian through a self-energy correction $H' = H_{\text{ssl}} + \Sigma_{\text{dis}}$, where $H_{\text{ssl}}$ is the spinful square-lattice Hamiltonian [Eq.~\ref{eq_hssl}]. 
The real part of $\Sigma_{\text{dis}}$ renormalizes the chemical potential as $\mu \rightarrow \mu + \text{Re}(\Sigma_{\text{dis}})$, effectively compensating the gate voltage and driving the system into a regime with open Fermi surfaces. 
As a result, highly disordered samples exhibit stable, nearly perfect spin-up polarization that persists even for long junction lengths [see Fig.~\ref{figureR3}(a-iv)]. 

In contrast, for the strong altermagnet [Fig.~\ref{figureR3}(b)], the polarization remains nearly perfect ($P \approx \pm 1$) across all disorder strengths [Fig.~\ref{figureR3}(b-i)].
The spin-resolved conductances clearly show that only one spin channel is active within a given gate-voltage window [Fig.~\ref{figureR3}(b-ii) and (b-iii)], reflecting the presence of open Fermi surfaces at all relevant energies. 
Importantly, the spin polarization remains robust and stable over long junction lengths [Fig.~\ref{figureR3}(b-iv)]. 

In summary, weak altermagnets with closed Fermi surfaces are sensitive to disorder; however, strong disorder can induce effective open Fermi surfaces, enabling fully spin-polarized transport. 
Strong altermagnets, by contrast, intrinsically host open Fermi surfaces and therefore maintain robust spin filtering even in the presence of strong disorder.

\subsection{two-sublattice tetragonal model}
A minimal two-dimensional model for altermagnetism is the two-sublattice tetragonal model \cite{Roig2024Minimal}, described in the basis 
$C_{\bm{k}}=(C_{\bm{k},a,\uparrow},C_{\bm{k},a,\downarrow},C_{\bm{k},b,\uparrow},C_{\bm{k},b,\downarrow})^{T}$ by  
\begin{eqnarray}
H_{\text{2t}}(\bm{k}) &=& t_{1}(\cos k_{x}+\cos k_{y}) + t_{2}\cos k_{x}\cos k_{y} - \mu \nonumber \\
&& + t_{3}\cos\frac{k_{x}}{2}\cos\frac{k_{y}}{2}\,\tau_{x} \label{eq_h2t} \\
&& + t_{4}\sin k_{x}\sin k_{y}\,\tau_{z} + t_{4}^{\prime}(\cos k_{x}-\cos k_{y})\,\tau_{z} \nonumber \\
&& + J\tau_{z}\sigma_{z}, \nonumber
\end{eqnarray}
where $C_{\bm{k},l,\sigma}$ annihilates an electron with momentum $\bm{k}$ and spin $\sigma$ on sublattice $l=a,b$. 
The parameters $\{t_{1},t_{2},t_{3},t_{4},t_{4}^{\prime},J\}$ can be fitted to density-functional theory calculations [cite]. 
The $d$-wave character of the altermagnetism originates from the interplay of $J$, $t_{4}$, and $t_{4}^{\prime}$: 
for finite $J$ and $t_{4}$ ($t_{4}^{\prime}$), Eq.\,(\ref{eq_h2t}) effectively describes a $d_{xy}$-wave ($d_{x^{2}-y^{2}}$-wave) altermagnet when $t_{4}^{\prime}=0$ ($t_{4}=0$). 

As a minimal model, Eq.\,(\ref{eq_h2t}) captures essential features of real materials. 
For example, with the parameters: $t_1=-0.1$, $t_2=0.1$, $t_3=1.7$, $t_4=0$, $t_4^{\prime}$, $\mu=0$ and $J_z=0.2$ guided by the DFT calculations \cite{Roig2024Minimal}, Hamiltonian [Eq.\,(\ref{eq_h2t})] describes the two-dimensional altermagnetic candidate RuO$_{2}$ \cite{Feng2022}, which exhibits a closed equal-energy surface at low energies \cite{He2025Evidence} and an open surface at higher energies \cite{Fedchenko2024Observation}. 
These features are shown in Fig.\,\ref{figureR1}(c). 
Moreover, by tuning the parameter $t_{3}$, Eq.\,(\ref{eq_h2t}) also reproduces the open equal-energy surfaces reported in strong altermagnetic MnTe \cite{Reichlova2024,Rial2024}, La$_{2}$O$_{3}$Mn$_{2}$Se$_{2}$, and Ba$_{2}$CaOsO$_{6}$ \cite{Ubiergo2025Atomic}, see Fig.\,\ref{figureR1}(d). 
Additionally, saddle points with high DOS act as transition markers between closed and open equal-energy surfaces in the weak regime, and between open surfaces with opposite spin polarization in the strong regime, consistent with the two-band model of Eq.\,(\ref{eq_hssl}) and Fig.\,\ref{figureR1}(a,b). 

Following the same procedure as in Eqs.\,(\ref{eq_hssltb})–(\ref{eq_P}), we compute the conductance and spin polarization of an altermagnetic junction described by Eq.\,(\ref{eq_h2t}), whose essential physics remains, although a more realistic model is employed.
The results, shown in Fig.\,\ref{figureR2}(c,d), again confirm the main conclusion:
As long as an open Fermi surface exists, perfect and stable spin polarization can be realized in altermagnetic junctions in both weak and strong regimes. 

\subsection{Six-band Lieb lattice model}
Another minimal two-dimensional altermagnetic model is the six-band Lieb lattice, which satisfies the symmetry constraints of $d$-wave altermagnetism \cite{Brekke2023Two}. 
The corresponding tight-binding Hamiltonian reads
\begin{equation}
H_{\text{6L}}^{\text{tb}} = 
t \sum_{\langle i,j \rangle,\sigma} c_{i,\sigma}^{\dagger}c_{j,\sigma}
+ t_{2} \sum_{\langle\!\langle i,j \rangle\!\rangle,\sigma} c_{i,\sigma}^{\dagger}c_{j,\sigma}
- J \sum_{i,\sigma,\sigma'} c_{i,\sigma}^{\dagger} (\bm{S}_{i}\cdot \bm{\sigma}_{\sigma,\sigma'}) c_{i,\sigma'}
- \mu \sum_{i,\sigma} c_{i,\sigma}^{\dagger}c_{i,\sigma}
+ \epsilon_{nm} \sum_{i \in nm,\sigma} c_{i,\sigma}^{\dagger}c_{i,\sigma},
\label{eq_h6Ltb}
\end{equation}
where $c_{i,\sigma}^{\dagger}$ ($c_{i,\sigma}$) creates (annihilates) an electron with spin $\sigma$ at site $i$. 
Here, $t$ is the nearest-neighbor hopping, $t_{2}$ the next-nearest-neighbor hopping between magnetic sites with opposite magnetization, $\mu$ the chemical potential, and $\epsilon_{nm}$ the nonmagnetic site energy (set to zero on magnetic sites). 
The exchange coupling $J$ describes the on-site interaction between itinerant and localized spins, with $\bm{S}_{i}=(0,0,S_{i}^{z})$. 
As there is no spin–orbit coupling, the Hamiltonian is block-diagonal in spin. 

After a Fourier transformation, the momentum-space Hamiltonian is
\begin{equation}
H_{\text{6L}}^{\text{tb}}(\bm{k}) =
\begin{pmatrix}
-\sigma_{z}JS - \mu & 2t\cos (ak_{x}) & 4t_{2}\cos (ak_{x})\cos (ak_{y}) \\
2t\cos (ak_{x}) & \epsilon_{nm}-\mu & 2t\cos (ak_{y}) \\
4t_{2}\cos (ak_{x})\cos (ak_{y}) & 2t\cos (ak_{y}) & \sigma_{z}S - \mu
\end{pmatrix},
\label{eq_h6L}
\end{equation}
with dispersions shown in Fig.\,\ref{figureR1}(e). 

Distinct from the previous two models, where the weak–strong altermagnet transition depends on parameter choices, the dispersion of Eq.\,(\ref{eq_h6L}) inherently contains both weak and strong altermagnetic features at different energies. 
For states near the upper and lower bands, the equal-energy surfaces resemble those of weak altermagnets [Fig.\,\ref{figureR1}(e-iii,e-iv) and (e-vii,e-viii)], whereas the middle two bands display strong-altermagnet behavior: two spin-polarized bands with open equal-energy surfaces along $k_{y}=0$, separated by a saddle point [Fig.\,\ref{figureR1}(e-v,e-vi)]. 

This dispersion directly governs the gate-tunable conductance and spin polarization of the junction [Fig.\,\ref{figureR2}(e)]. 
At low and high $U_{g}$, the system behaves like a weak altermagnet, exhibiting a transition from unstable to stable perfect spin polarization. 
At intermediate $U_{g}$, however, the strong-altermagnet character dominates, leading to a flip of the perfectly spin-polarized conductance. 
Thus, the Lieb lattice illustrates how weak and strong altermagnetic transport characteristics can naturally coexist in a single model, depending on energy or gate voltage.